\documentclass{aa}
\usepackage[varg]{txfonts}
\usepackage{natbib}
\usepackage{graphicx}
\usepackage{times}
\usepackage{color}
\bibpunct{(}{)}{;}{a}{}{,}
\newcommand{\va}{V1294~Aql }
\newcommand{\ve}{V1294~Aql}

\newcommand{\Mnom}{\hbox{$\mathcal{M}^{\mathrm N}_\odot$}}
\newcommand{\Rnom}{\hbox{$\mathcal{R}^{\mathrm N}_\odot$}}

\newcommand{\spefo}{{\tt SPEFO} }

\newcommand{\respefoe}{{\tt reSPEFO}}

\newcommand{\fotel}{{\tt FOTEL} }

\newcommand{\tria}{\hbox{$\bigtriangleup$}}
\newcommand{\ubv}{\hbox{$U\!B{}V$}}
\newcommand{\bvr}{\hbox{$B{}VR$}}
\newcommand{\bv}{\hbox{$B\!-\!V$}}
\newcommand{\ub}{\hbox{$U\!-\!B$}}
\newcommand{\vr}{\hbox{$V\!-\!R$}}
\newcommand{\ubvr}{\hbox{$U\!B{}V\!R$}}
\newcommand{\uvby}{\hbox{$uvby$}}
\newcommand{\hp}{\hbox{$H_{\rm p}$}}

\newcommand{\p}{$\pm$}

\newcommand{\arcm}{$^\prime$}

\newcommand{\m}{$^{\rm m}\!\!.$}

\newcommand{\D}{$^{\rm d}\!\!.$}

\newcommand{\ks}{km~s$^{-1}$}
\newcommand{\vsin}{$v$~sin~$i$ }

\newcommand{\tef}{$T_{\rm eff}$ }
\newcommand{\lgg}{{\rm log}~$g$ }
\newcommand{\ms}{M$_{\odot}$}

\newcommand{\ha}{H$\alpha$ }

\newcommand{\hae}{H$\alpha$}

\newcommand{\he}{\ion{He}{i}~6678 }
\newcommand{\hea}{\ion{He}{i}~6678}

\newcommand{\ii}{\hbox{\'{\i}}}

\begin{document}

   \title{V1294~Aql = HD 184279: A bad boy among Be stars\\
 or an~important clue to the Be phenomenon?
\thanks{Based on new spectroscopic observations
from the ESO FEROS echelle spectrograph;
CCD coud\'e spectrograph of the Astronomical Institute of Czech
Academy of Sciences in Ond\v{r}ejov; CCD coud\'e
spectrograph of the Dominion Astrophysical Observatory;
echelle BESO spectrograph of the Universit\"atssternwarte
Bochum; CCD spectrographs of Castanet-Tolosan Observatory; CHIRON
echelle spectrograph of Cerro Tololo Observatory; and amateur
spectra from the BeSS database, and on photometric observations from
Hvar, Toronto, San Pedro M\'artir, Tubitak and \c{C}anakkale Observatories,
and $H_{\rm p}$ photometry from the ESA Hipparcos mission.}
\fnmsep\thanks{Tables 3, 4, and 5 are available only in electronic form
 at the CDS via anonymous ftp to cdarc.u-strasbg.fr (130.79.128.5)
 or via http://cdsweb.u-strasbg.fr/cgi-bin/qcat?J/A+A/}
}
%\subtitle{Duplicity and remarkable spectral variations of
%          V1294~Aql = HD 184279}
  \author{P.~Harmanec\inst{1}\and
          H.~Bo\v{z}i\'c\inst{2}\and
          P.~Koubsk\'y\inst{3}\and
          S.~Yang\inst{4}\and
          D.~Ru\v{z}djak\inst{2}\and
          D.~Sudar\inst{2}\and
          M.~\v{S}lechta\inst{3}\and
          M.~Wolf\inst{1}\and
          D.~Kor\v{c}\'akov\'a\inst{1}\and
          P.~Zasche\inst{1}\and
          A.~Opli\v{s}tilov\'a\inst{1}\and
          D.~Vr\v{s}nak\inst{5}\and
          H.~Ak\inst{6}\and
          P.~Eenens\inst{7}\and
          H.~Baki\c{s}\inst{8}\and
          V.~Baki\c{s}\inst{8}\and
          S.~Otero\inst{9}\and
          R.~Chini\inst{10,11}\and
          T.~Demsky\inst{10}\and
          B.N.~Barlow\inst{12}\and
          P.~Svoboda\inst{13}\and
          J.~Jon\'ak\inst{1}\and
          K.~Vitovsk\'y\inst{1}\and
          A.~Harmanec\inst{14}
}
%  \thanks{Guest investigator, Dominion Astrophysical Observatory,
%          Victoria, BC, Canada}

   \offprints{P. Harmanec\,\\
               \email Petr.Harmanec@mff.cuni.cz}

  \institute{
   Astronomical Institute of Charles University,
   Faculty of Mathematics and Physics,\hfill\break
   V~Hole\v{s}ovi\v{c}k\'ach~2, CZ-180 00 Praha~8 - Troja, Czech Republic
 \and
  Hvar Observatory, Faculty of Geodesy, University of Zagreb,
  Ka\v{c}i\'ceva~26, 10000~Zagreb, Croatia
 \and
   Astronomical Institute, Czech Academy of Sciences,
   CZ-251 65 Ond\v{r}ejov, Czech Republic
 \and
   Physics \& Astronomy Department, University of Victoria,
  PO Box 3055 STN CSC, Victoria, BC, V8W 3P6, Canada
\and
   Faculty of Electrical Engineering and Computing,
   University of Zagreb, Unska~3, 10000~Zagreb, Croatia
\and
   Astronomy and Space Sci. Dept., Science Faculty,
   Erciyes University, 38039, Kayseri, Turkiye
\and
   Dept. of Astronomy, University of Guanajuato,
   36000 Guanajuato, GTO, Mexico
\and
    Dept. of Space Sciences and Technology, Akdeniz University,
   07058, Antalya, Turkiye
\and
   Buenos Aires, Argentina and American Association of Variable Star
   Observers (AAVSO), \hfill\break
   49 Bay State Road, Cambridge, MA 02138, USA
\and
   Astronomisches Institut,
   Ruhr--Universit\"at Bochum,
   Universit\"atsstr. 150,
   44801 Bochum, Germany
\and
   Instituto de Astronom{\'i}a,
   Universidad Cat\'{o}lica del Norte,
   Avenida Angamos 0610,
   Casilla 1280
   Antofagasta, Chile
\and
   Department of Physics, High Point University,
   One University Way, High Point, NC 27268, USA
\and
   Private Observatory, Vypustky 5, Brno, CZ-614 00 Czech Republic
\and
   Faculty of Mathematics and Physics, Charles University, Prague, Czech Republic
}
\date{Received \today}

  \abstract{
A reliable determination of the basic physical properties and variability
patterns of hot emission-line stars is important for understanding
the Be phenomenon and ultimately, the evolutionary stage of Be stars.
This study is devoted to one of the most remarkable Be stars, V1294~Aql = HD~184279.
We collected and analysed spectroscopic and photometric observations covering
a~time interval of about 25000~d (68~yr). We present evidence that
the object is a single-line 192.9~d spectroscopic binary and estimate that the secondary
probably is a hot compact object with a mass of about 1.1-1.2~\ms. We found and
documented very complicated orbital and long-term spectral, light, and colour variations,
which must arise from a~combination of several distinct variability patterns.
Attempts at modelling them are planned for a follow-up study. We place the
time behaviour of \va into context with variations known for some other
systematically studied Be stars and discuss the current ideas about
the nature of the Be phenomenon.}


\authorrunning{P. Harmanec et al.}
\titlerunning{Duplicity and variability of the Be star \ve}
\maketitle

\section {Introduction}
The B0.5IV star \va (also known as HD 184279, BD+03$^\circ$4065, HIP~96196, SAO 124788,
MWC 319; $\alpha_{2000.0}$ = 19$^{\rm h}$33$^{\rm m}$36.$\!\!^{\rm s}$9191,
$\delta_{2000.0}$ = +03$^\circ$45\arcm40\farcs779) is one of a few Be stars
for which photometric variations over a few decades are rather well documented.
This is so thanks to the curious fact that the star was recommended
as a secondary standard for \ubv\ photometry by \citet{john54} and
\citet{john55} and therefore was relatively often observed by various
observers. When \citet{dahn73} and \citet{temp76} reported pronounced
light variations of this `secondary standard', other research teams reanalysed
their photometric observations to document the variations further.
This early history has been summarised in detail by \citet{zarf11}, who
confirmed an earlier suspicion by \citet{balle82} that a correlation exists 
between light and spectral changes. They collected the records about the presence 
of Balmer emission and compared them to the light and colour variations based 
on historical records and based also on their new \ubv\ photometry from Hvar 
Observatory. They noted that the observed cyclic long-term
variability contradicts what is usually observed for the long-term
$V/R$ changes of the Balmer emission lines and expressed doubts about
their interpretation by the model of a revolving elongated disk.
Comparing the light curve of \va to similar
light curves of BU~Tau and V744~Her, they noted that the development of a new
shell phase in all three objects was accompanied by a pronounced light
decrease. This type of variability was later classified as an inverse
correlation between the brightness and emission strength by \citet{hvar83}
and was interpreted as an essentially geometrical effect for situations in which
the circumstellar envelope in question is seen more or less equator-on.
A~study of the disk models, which grow in size and/or density
with time, by \citet{sigut2013} basically confirmed this conjecture.
The correlation was documented by \citet{horn83} throughout the entire interval
of 50 years covered by spectral records with the help of a~long series of
photographic magnitudes.

\begin{table}
\begin{center}
\caption[]{Journal of electronic spectra.}\label{jourv}
\begin{tabular}{ccrcrl}
\hline\hline\noalign{\smallskip}
Spg.&Time interval&No.   &Wavelength &Spectral  \\
 No.&             &of    & range     &res.\\
    &(HJD-2450000)&RVs   & (\AA)   \\
\noalign{\smallskip}\hline\noalign{\smallskip}
 1&1370.4          & 1&6520--6600& 2500\\
 2&2420.53--2778.67& 2&6340--6860& 6000\\
 3&2860.40--2870.48& 3&6530--6695& 6000\\
 3&2918.34         & 1&6520--6645& 6000\\
 4&4351.34         & 1&6470--6730& 6000\\
 5&4653.48         & 1&4750--6990&10000\\
 6&4944.64         & 1&4270--6910&10000\\
 7&2461.85         & 1&3770--9220&48000\\
 8&2749.62--6152.49&58&6255--6767&12700\\
 9&6475.49--9483.31&66&6260--6735&12700\\
10&2771.92--4443.60&21&6150--6760&21700\\
10&9070.86--9499.64& 4&6325--6930&21700\\
11&5057.59--6890.56& 9&3800--8750&40000\\
12&6105.50--9480.39& 8& various  &$>10000$\\
13&9489.51--9526.50& 3&4500--8900&79000\\
\noalign{\smallskip}\hline\noalign{\smallskip}
\end{tabular}
\tablefoot{Column ``Spg. No.": \ \
1...Castanet-Tolosan T190 instrument, observer C. Buil -- see
\url{http://www.astrosurf.com/buil/us/bestar.htm};
2... Castanet-Tolosan T212;
3... Castanet-Tolosan R128;
4... Castanet-Tolosan T280+Lhires III;
5... Castanet-Tolosan C11+eShel+Audine KAF1600;
6... Castanet-Tolosan C11+eShel+QSI532;
7... ESO 1.52 m reflector, FEROS spectrograph;
8... Ond\v{r}ejov 2.0 m reflector, coud\'e grating spectrograph,
    CCD SITe5 2030x800
    pixel detector;
9... Ond\v{r}ejov 2.0 m reflector, coud\'e grating spectrograph,
    CCD Pylon Excelon 2048x512 pixel detector;
10... DAO 1.22 m reflector, coud\'e grating McKellar spectrograph,
     CCD Site4 detector;
11... HTP 1.5~m reflector, BESO echelle spectrograph;
12... Amateur spectra from the BeSS database
     (\url{http://basebe.obspm.fr/basebe});
13... Cerro Tololo 1.5 m reflector, CHIRON echelle spectrograph.
}
\end{center}
\end{table}

 The first detailed spectroscopic study of \va was published in two
papers by \citet{balle87} and \citet{balle89}. They demonstrated the presence
of cyclic long-term $V/R$ variations of the double Balmer emission lines
with a cycle length of some 6~yr and concluded that they are caused by
a slow revolution of an elongated envelope
\citep[the model of][]{mclaughlin61b,mclaughlin61a}.
They also investigated the possibility whether the star might be
a spectroscopic binary, but found no evidence for it.
\citet{mou98} studied the correlation of the second Balmer jump with visual
brightness and long-term radial-velocity (RV) changes for a number of Be stars. 
For \ve, they concluded that the Balmer jump parameter $D$\footnote{For normal stars,
the Balmer jump parameter D, which is defined as $\log$ of the flux shortward
3700~\AA\  divided by the flux shortward 3700~\AA, is a good measure of
the stellar effective temperature. The presence of circumstellar matter
can cause the second Balmer jump, which is occasionally seen in emission.}
correlated with both these
quantities, attaining maximum in 1979, when the long-term cyclic RV
variations were also at maximum.

\begin{table*}
\caption[]{Journal of available photometry with known dates of observations.}
\label{jouphot}
\begin{flushleft}
\begin{tabular}{rcrccll}
\hline\noalign{\smallskip}
Station&Time interval& No. of &Passbands&HD of comparison&Source\\
       &(HJD$-$2400000)&obs.  &  &/ check star\\
\noalign{\smallskip}\hline
\hline\noalign{\smallskip}
44&36431.77--36462.78& 42& $V$& 184663  & \citet{lynds59}; see the text\\
59&36795 -- 41544$^{**}$&3&\ubv&all-sky & \citet{cousins63,cousins73}\\
34&  39370 \p 30$^*$ &  1&\ubv& all-sky & \citet{moreno71}\\
90&39928 -- 42988$^*$& 40&$BV$& all-sky & \citet{temp76}  \\
26&40449.4 -- 40452.4$^*$&  2&\ubv& 183227  & \citet{haupt74} \& priv.com.\\
92&    40453.5$^*$   &  1&\ubv& all-sky & \citet{wal72}   \\
91&    41230.7$^*$   &  1&\ubv& all-sky & \citet{dahn73}  \\
37&42894.92--43391.55&  3&Geneva&all-sky& Burki (priv.com.)\\
30&43382.85--43690.85&  6&13C & all-sky & \citet{alvschu82}\\
59&43401 -- 43796$^{**}$&2&$VRI$&all-sky& \citet{cousins78}\\
 1&44073.40--48128.39&406&\ubv& 183324 / 183227 & this paper\\
12&44028 --  44433$^*$ &  4&\uvby& all-sky & \citet{kozok85})\\
20&45099.81--49951.73& 90&$BV$& 183324 / 183227 & this paper\\
12&45249.56--49591.69&121&\uvby&183324 / 183227 & \citet{esovar91,esovar93}\\
30&45451.00--45451.98&  2&13C & all-sky & \citet{schugui84}\\
 4&45525.50--45573.38&  4&\ubv& 183227 / 184663 & this paper\\
61&47963.79--49051.86& 90&$V$ & all-sky         &\citet{esa97}\\
93&51979.90--55137.52&795&$V$ & all-sky         &\citet{pojm2002}\\
30&52060.81--52752.94& 45&\ubv& 183324 / 183227 & this paper\\
 1&52076.49--56475.46&624&\ubv& 183227 / 184663 & this paper\\
66&52763.49--52764.54& 10&\ubv& 183324 / 183227 & this paper\\
89&52850.40--53309.29& 69&\ubv& 183324 / 183227 & this paper\\
 1&56488.41--59483.27&590&\ubvr&183227 / 184663 & this paper\\
 2&57297.27--58089.25&533/555/549&\bvr &183227 / 184663 & this paper\\
\noalign{\smallskip}\hline
\end{tabular}
\tablefoot{$^*$ Inaccurate Julian dates; $^{**}$ Julian dates uncertain by $\pm15$~d\\
In the column {\sl ``Station",} the individual observing stations are
identified by the running numbers from the Praha/Zagreb photometric archives:\\
1... Hvar 0.65~m reflector, EMI tubes;
2... Brno private observatory of P.~Svoboda, Sonnar 0.034~m refractor with a CCD camera;
4... Ond\v{r}ejov 0.65~m reflector, EMI tube;
12... La Silla 0.50~m Danish reflector: Part 1 are all-sky observations, the rest
      are differential observations secured during a~long-term campaign. The original \uvby\
      observations were transformed to \ubv;
20... Toronto 0.40~m reflector, EMI 6094 tube;
26... Haute Provence 0.60~m reflector, Lallemand tube;
30... San Pedro M\'artir 0.84~m and 1.50~m reflectors (13C photometry),
      more recent \ubv\ observations were secured with the 0.84 m
      reflector and Cuenta-pulsos photometer;
34... Lick 0.61 m Boller \& Chivens telescope, refrigerated S20 tube;
37... Jungfraujoch Sphinx mountain station 0.40 m reflector, Geneva photometer;
44... MtPalomar 0.51~m reflector, EMI~6094 tube;
59... Cape Town 1.0~m reflector, \ubv\ photometer;
61... Hipparcos \hp\ photometry transformed into Johnson $V$;
66... T\"ubitak National Observatory 0.40~m reflector with SSP-5A photometer;
89... \c{C}anakkale Mountain Station 0.40~m reflector, SSP-5 photometer;
90... Teramo 0.40~m refractor;
91... US Naval Observatory Ritchey-Chretien 1.0~m reflector, EMI 9524B tube;
92... Stefanion ESO Netherlands Van Straaten 0.40~m reflector, EMI 6256B tube;
93... ASAS3 automatic all-sky photometric survey.
}
\end{flushleft}
\end{table*}

\citet{mennick97} discussed possible observational tests of the models
of long-term cyclic variations of the Be star disks. Investigating
spectral, light, and colour variations of \va over the time interval
from JD~2440000 to 2450000, they noted that the brightness extrema
corresponded to the phase transitions of the $V/R$ variations from
$V/R>1$ to $V/R<1$ and vice versa, but admitted that this behaviour was
not seen during the third $V/R$ transition over the investigated
interval of time. They tentatively concluded, however, that the disk
of \va is seen more or less equator-on \citep[in agreement with][]{zarf11}
and that the revolution
of its elongated structure (``one-armed global oscillation") projected
against the star during the phase transitions causes the observed
light decreases through attenuation of the stellar flux.

\citet{chini2012} carried out a spectroscopic survey of 249 O-type
and 581 B-type stars in a search for the duplicity. Based on variable RV,
they reported that \va is a single-line spectroscopic binary.
In a~search for hot subdwarf companions to Be stars in the spectra
obtained by the International Ultraviolet Explorer (IUE), \citet{wang2018}
reported a~null detection for such a companion to \ve.
\citet{brandt2021} cross-corelated the Hipparcos and Gaia catalogues in
an effort to identify astrometrically accelerating objects. \va was
identified as a candidate for a~system with a~faint companion.

%-------------------------------------------------------------------------------
\section{Observations and reductions}
\subsection{Spectroscopy}
New electronic spectra were obtained at four observatories: with
the Ond\v{r}ejov 2~m reflector (OND) and a~coud\'e
spectrograph, with the Dominion Astrophysical Observatory 1.22~m reflector (DAO)
and a coud\'e spectrograph, with the Cerro Armazones 1.5~m Hexapod Telescope (HPT),
the Bochum Echelle Spectroscopic Observer (BESO)
spectrograph, which is similar to FEROS \citep{beso}, and with the Cerro Tololo Inter-American
Observatory (CTIO) 1.5~m reflector with the CHIRON echelle spectrograph
\citep{toko2013}.
We also used one archival ESO FEROS echelle spectrum \citep{feros,feros2},
the medium-resolution CCD  spectra obtained and published by Christian Buil,
\footnote{For the description of his instrumentation and data reduction, see
\url{http://www.astrosurf.com/buil/us/bestar.htm}\,.} and a selection of
amateur CCD spectra with resolutions better than 10000 from the BeSS
spectroscopic database \citep{neiner2011}.
Table~\ref{jourv} lists a journal of all spectral observations used.

The initial reduction of all Ond\v{r}ejov and DAO spectra (bias subtraction,
flat-fielding, creation of 1D spectra, and wavelength calibration) was carried
out in {\tt IRAF}. Initial reduction of the HTP and CTIO spectra
was carried out at the respective observatories. Rectification, removal
of residual cosmics and flaws, and RV measurements of all spectra were carried
out with the Pascal program \spefo \citep{sef0,spefo}, namely the latest
version 2.63 developed by J.~Krpata \citep{spefo3}.
\spefo displays direct and flipped traces of the line
profiles superimposed on the computer screen that the user can slide
to achieve a precise overlapping of the parts of the profile for which the RV
is to be measured.
All RVs measurements were carried out independently by PH and also by AH, who
studied the spectra as a~part of his student's research project. In addition to the wings of the \ha and \he emission, we also measured the bottom of the
(sometimes asymmetric) cores of the \hae, \hea, \ion{Si}{ii}, and \ion{Fe}{ii}
shell absorptions. Both sets of measurements were
intercompared, the larger deviations were carefully
checked, and the mean of the measurements for each line was then used here.
Only after these RV measurements were completed was a new program
for spectral reduction \respefoe, a~modern replacement of \spefo, written
in JAVA and running on different platforms (Linux, Windows) developed
by A.~Harmanec.\footnote{\url{https://astro.troja.mff.cuni.cz/projects/respefo}}
It can, among other things, import the spectra that were originally reduced in \spefo
and treat spectra stored as FITS files.
We used this new program to measure line intensities, equivalent widths, and
the $V/R$ ratio of the double \ha emission.\\

\subsection{Photometry}
We attempted to collect all available observations with known
dates of observations. Basic information about all data sets
can be found in Table~\ref{jouphot}, and the details of the photometric
reductions and standardisation are described in Appendix~\ref{apb}.

For the convenience of other investigators, we also publish all our individual
observations together with their HJDs. All measured RVs are listed in Table~3,
spectrophotometric quantities are provided in Table~4, and photometric observations
are collected in Table~5.  These three tables are available in electronic form only.

\section{Long-term spectral, light, and colour changes of \ve}
As mentioned above, we attempted to collect and homogenise all available
photometric observations and measurements of RVs and line strengths from
the records with known dates of observations. Long-term behaviour of these
quantities and their mutual correlations are discussed in the following
sub-sections.

\begin{figure}
\resizebox{\hsize}{!}{\includegraphics{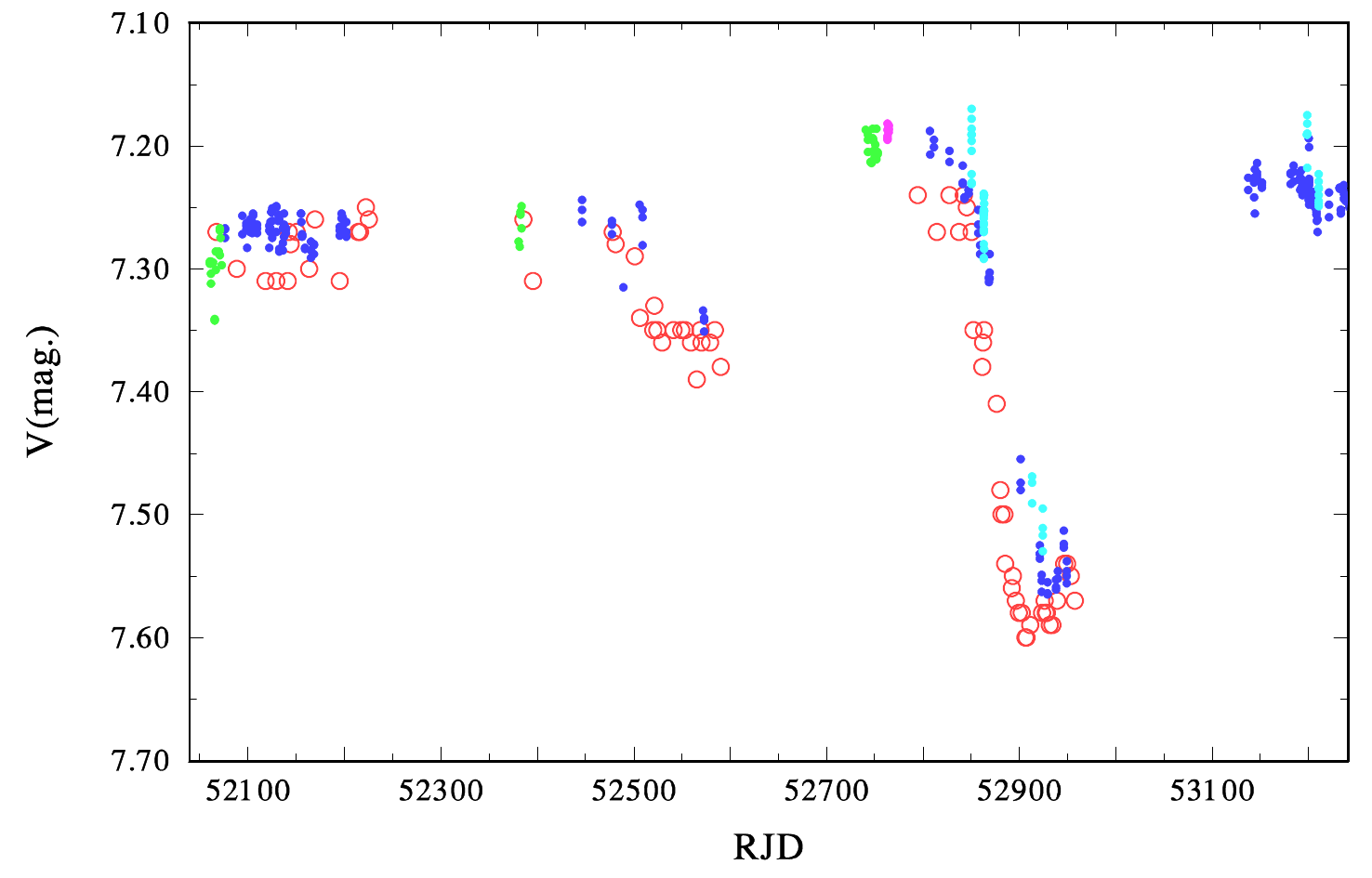}}
\caption{Time plot of the Otero visual brightness estimates
(empty red circles) along with Johnson $V$ photometry from Hvar (blue dots),
        SPM (green), \c{C}anakkale (cyan), and TUG (magenta).}
\label{vis-v}
\end{figure}

\subsection{Value of visual estimates of brightness}
The visual estimates of brightness by skilful amateur observers are commonly
used to determine the times of minima (or maxima) of periodic
variables and to monitor light changes of variables with a large
amplitude (over several magnitudes). The scatter band of
visual estimates is typically about 0\m1 to 0\m15, but it can be pushed down by
talented individuals who, moreover, follow a few principles:
(a) to perform only one visual estimate per night (without recalling the previous
one), and (b) to reduce the estimates to Johnson $V$ magnitudes of the
comparison stars (known to one thousandth of a magnitude), not to
the Harvard scale of magnitudes, which is only accurate to 0\m1.
One of us (SO) contacted the first author of this study back in 2003
to inform him that he had observed another light decrease of \va for
about 0\m3. We then agreed to test his ability to obtain accurate visual
estimates via parallel photoelectric observations at Hvar,
San Pedro M\'artir (SPM), Tubitak National Observatory
(TUG), and \c{C}anakkale. Figure~\ref{vis-v}
shows the comparison of visual estimates and Johnson $V$ photoelectric
photometry from several stations over the time interval covered by visual
estimates. The visual estimates agree very well with
the general trend of variations that were recorded via photoelectric photometry, but
the deep light minimum appears broader than that recorded by photoelectric
photometry. One possible reason is that the minimum was observed
close to the end of visibility of the star in the sky and the visual
estimates were not corrected for the differential extinction.

\subsection{Correlation between the long-term light and spectral changes
in time}

\begin{figure}
\resizebox{\hsize}{!}{\includegraphics[angle=-90]{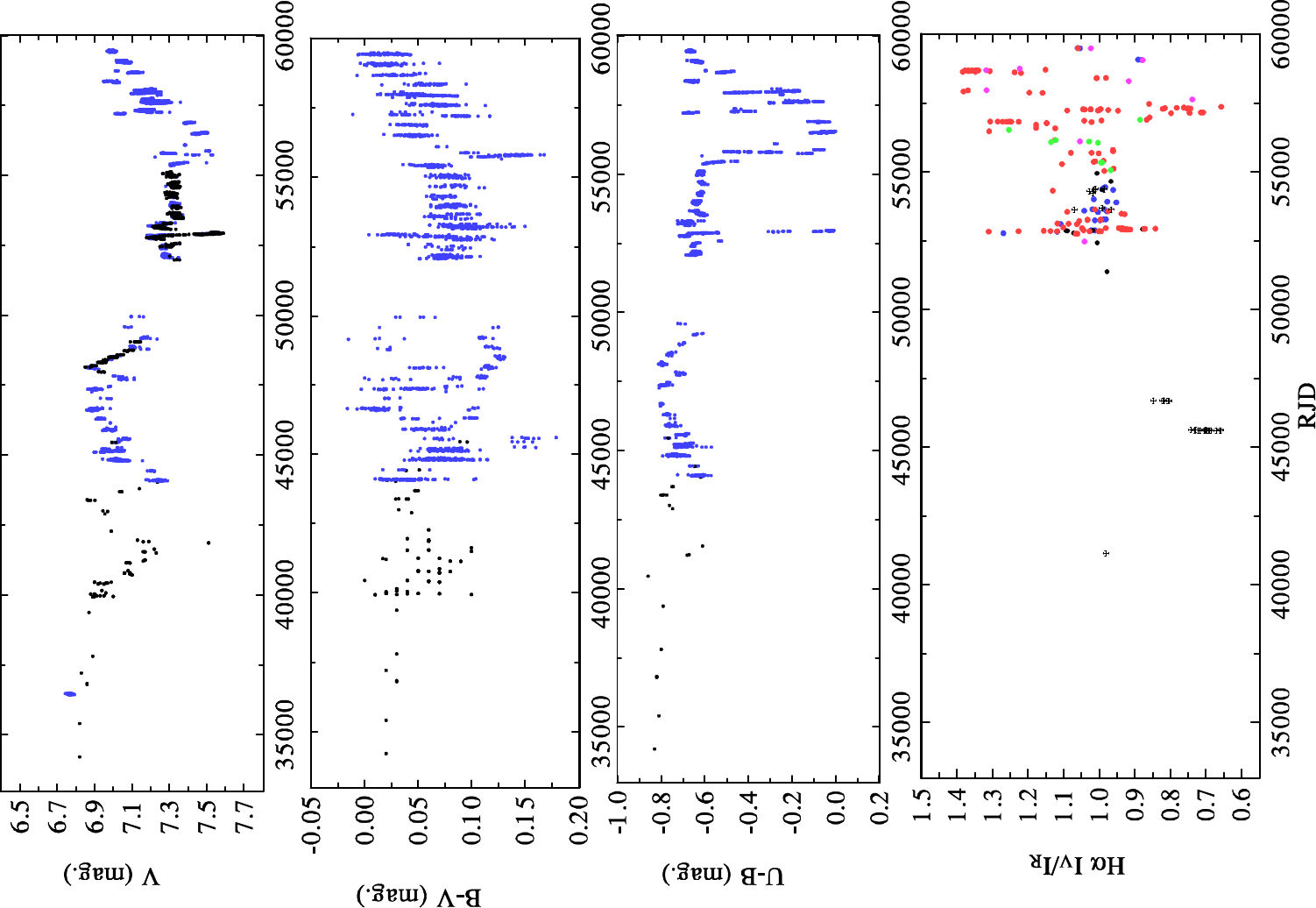}}
\caption{Time plot of available observations over the whole time interval
of about 25000~d covered by the data. Top panel: Yellow brightness observations,
which could be transformed into Johnson $V$ magnitude. Second and third panel:
Available \bv\  and \ub\ colour index observations. Bottom panel: $V/R$ changes in the peaks of the double \ha emission.
In the three panels with photometry, the differential observations
are shown as blue dots, and all-sky observations are shown as black dots. In the bottom panel,
blue circles denote the DAO spectra, red circles show the OND spectra, green circles represent BESO spectra,
magenta circles show BeSS spectra, black circles show Castanet Tolosan spectra, and the black crosses plot the data
from the literature.}
\label{all}
\end{figure}

\begin{figure}[t]
\resizebox{\hsize}{!}{\includegraphics[angle=-90]{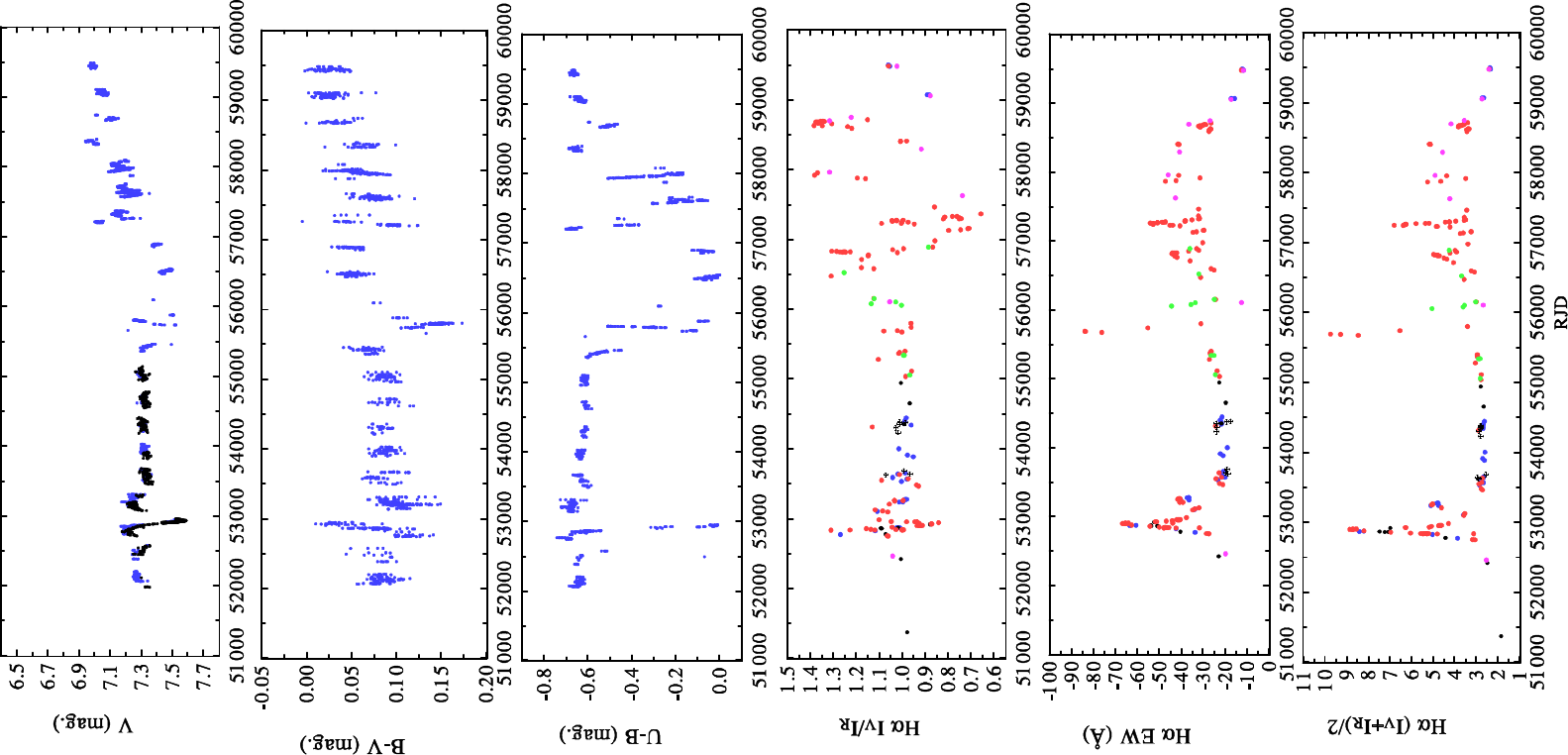}}
\caption{Time plot showing the correlation between the secular brightness
and colour variations in the $V$ -band magnitudes and \bv\ and \ub\ colour
indices (black shows all-sky and blue shows differential observations),
and the $V/R$ changes, EW, and strength of the
\ha emission for the more recent electronic spectra. The colour symbols for 
spectra from different sources are the same as in Fig. 2.}
\label{all-n}
\end{figure}

\begin{figure}[t]
\resizebox{\hsize}{!}{\includegraphics{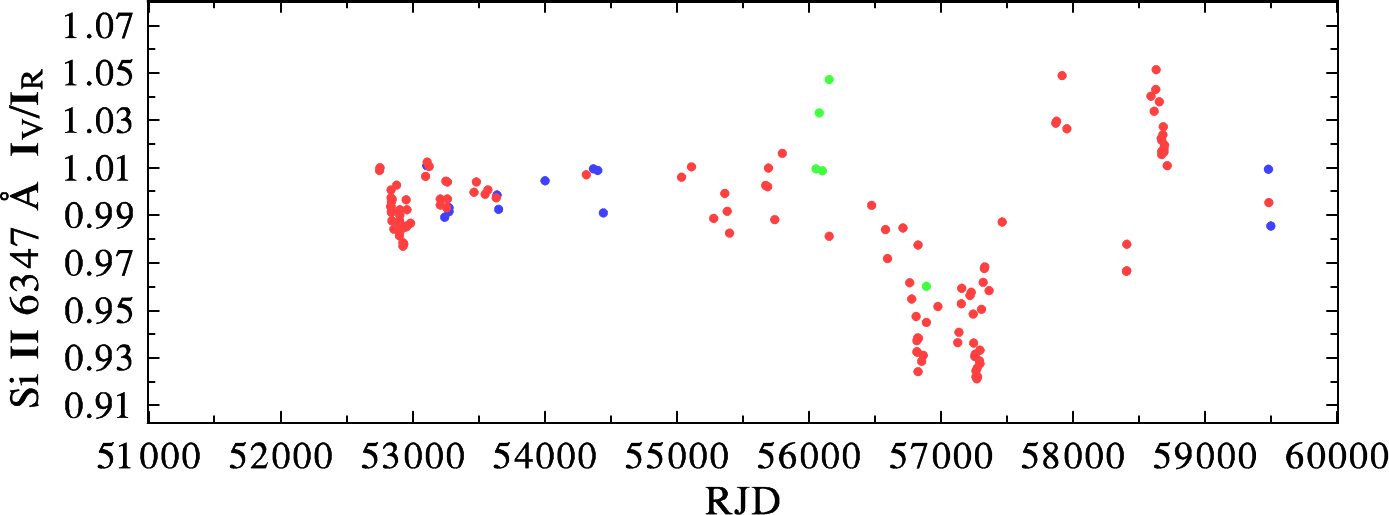}}
\caption{Time plot of the $V/R$ changes recorded for the electronic spectra
in the \ion{Si}{ii}~6347~\AA\ line. The colour symbols for spectra from different
sources are the same as in Fig.~\ref{all}.}
\label{si1vr}
\end{figure}

\begin{figure}[t]
\resizebox{\hsize}{!}{\includegraphics{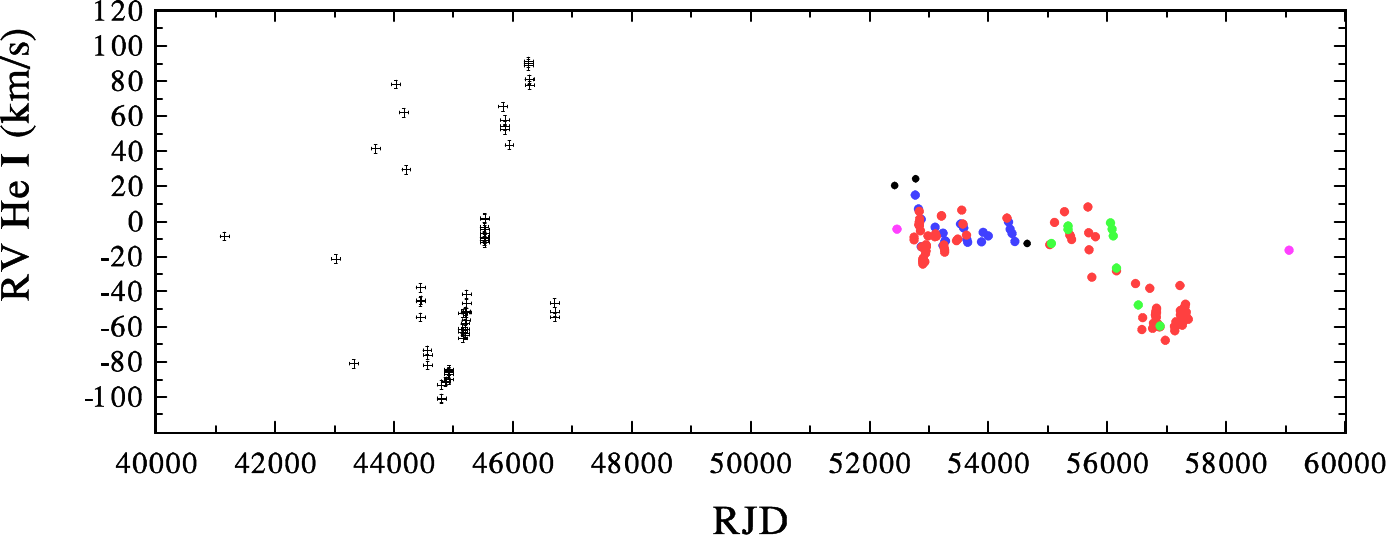}}
\resizebox{\hsize}{!}{\includegraphics{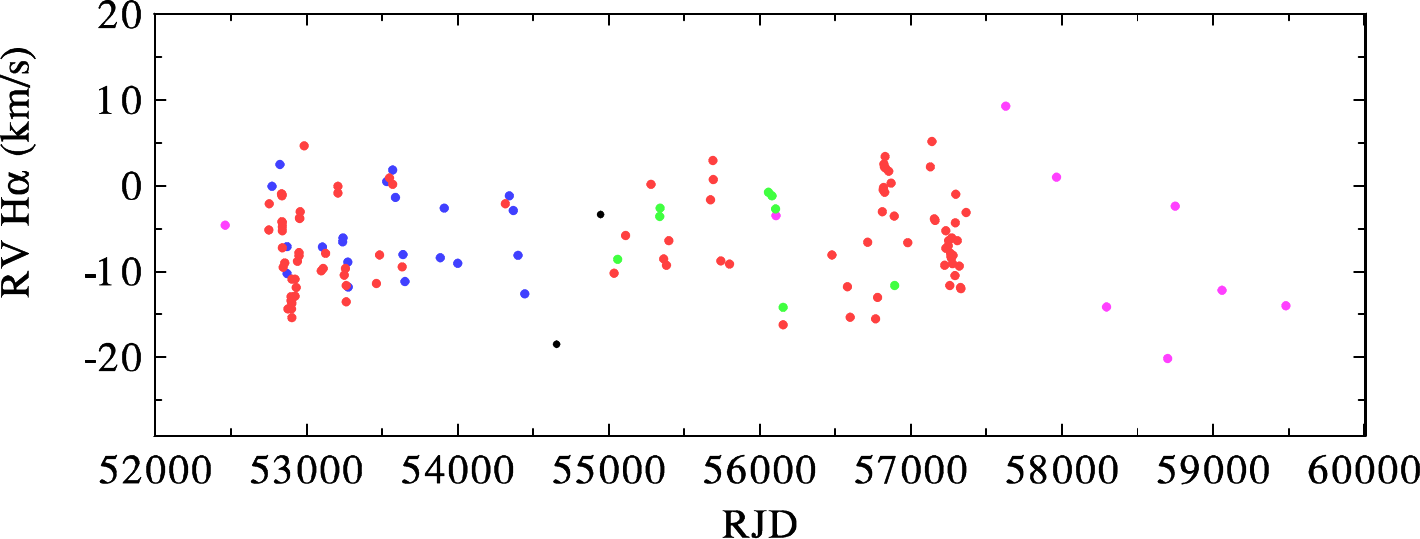}}
\caption{Time plot of radial velocities. Top panel: RV of the \ion{He}{i}
shell lines. Bottom panel: RV measured on the steep wings of the \ha emission.
Data from individual instruments are shown by different symbols:
the circles are the same as in Fig. 2, and
black crosses plot data from \citet{balle89}. The ranges on the two
axes in the two plots are different.}
\label{rvtime}
\end{figure}

Figure~\ref{all} is a time plot of all available observations secured in
or transformed into the Johnson \ubv\ magnitudes. The top panel
shows that the usual brightness level in $V$ is occasionally
disturbed by rapid light decreases of different durations.
Moreover, we note that there is also a secular, steady slow light decrease
of the undisturbed brightness of the star outside the more rapid light
decreases until about HJD~2455000, when it suddenly changed to
a~steeper secular light increase. We return to this new phenomenon
in a separate section below. The second panel of Fig.~\ref{all} shows
that the \bv\ index followed the brightness changes, but with a small amplitude,
while the \ub\ index showed a similar pattern of changes, but with a~larger
amplitude and with a~more or less steady secular reddening.

An enlarged Fig.~\ref{all-n} covers only the more recent time interval,
when electronic spectra became available. It shows that in the time intervals
that are sufficiently densely covered by the data, the sharp light decreases are
accompanied by similarly sharp strengthening in the \ha emission.

Figure~\ref{si1vr} shows the $V/R$ changes of \ion{Si}{ii}~6347~\AA\ line.
It reveals a~pattern similar to that observed for \hae, but the time coverage
is less dense because these variations could not be measured in time intervals
when the emission was faint.

Figure~\ref{rvtime} shows the variation of the shell absorption RVs
\citep[characterised by \ion{He}{i} RVs, for which data are also
published by][]{balle89} and emission-line RVs measured on the wings of
the \ha line in electronic spectra.  We note that while the shell RVs
shows large cyclic changes that have also been observed for a number of other Be stars,
the RVs measured on the wings of the \ha emission are secularly stable and
show only mild changes on a shorter timescale.

\subsection{Unusual colour variations}

\begin{figure*} [t]
\includegraphics[angle=0,scale=0.95]{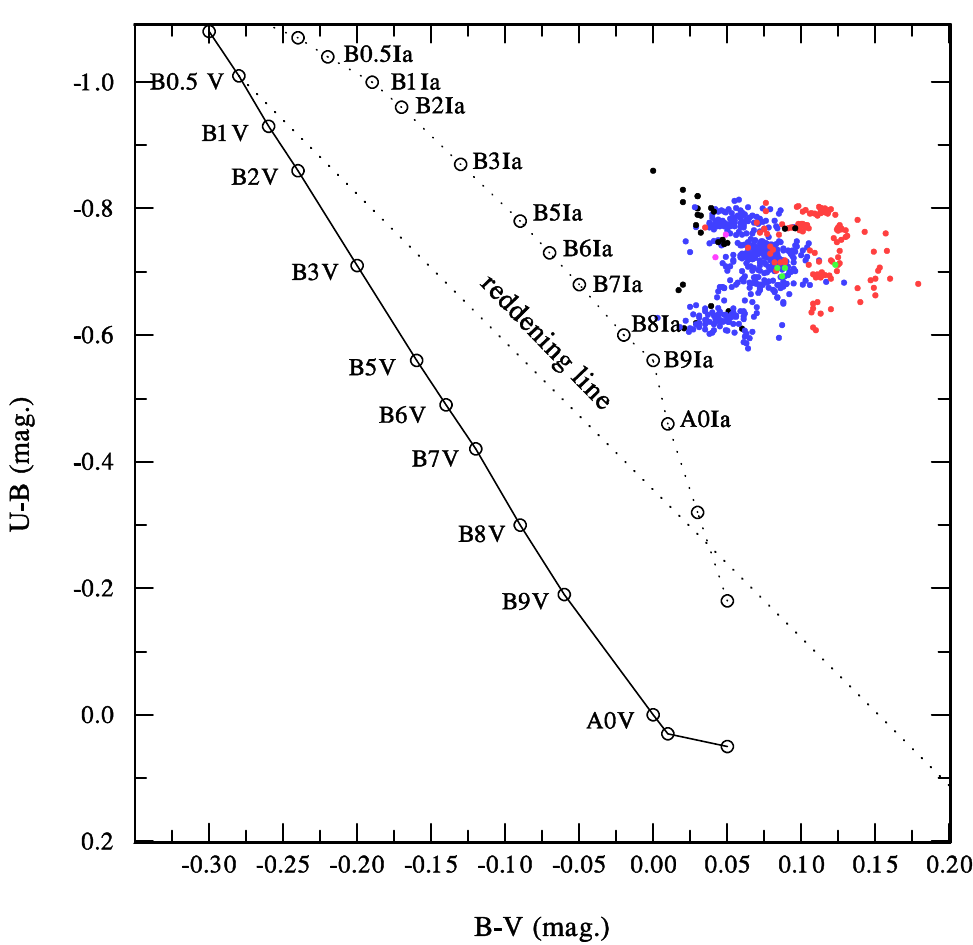}
\includegraphics[angle=0,scale=0.95]{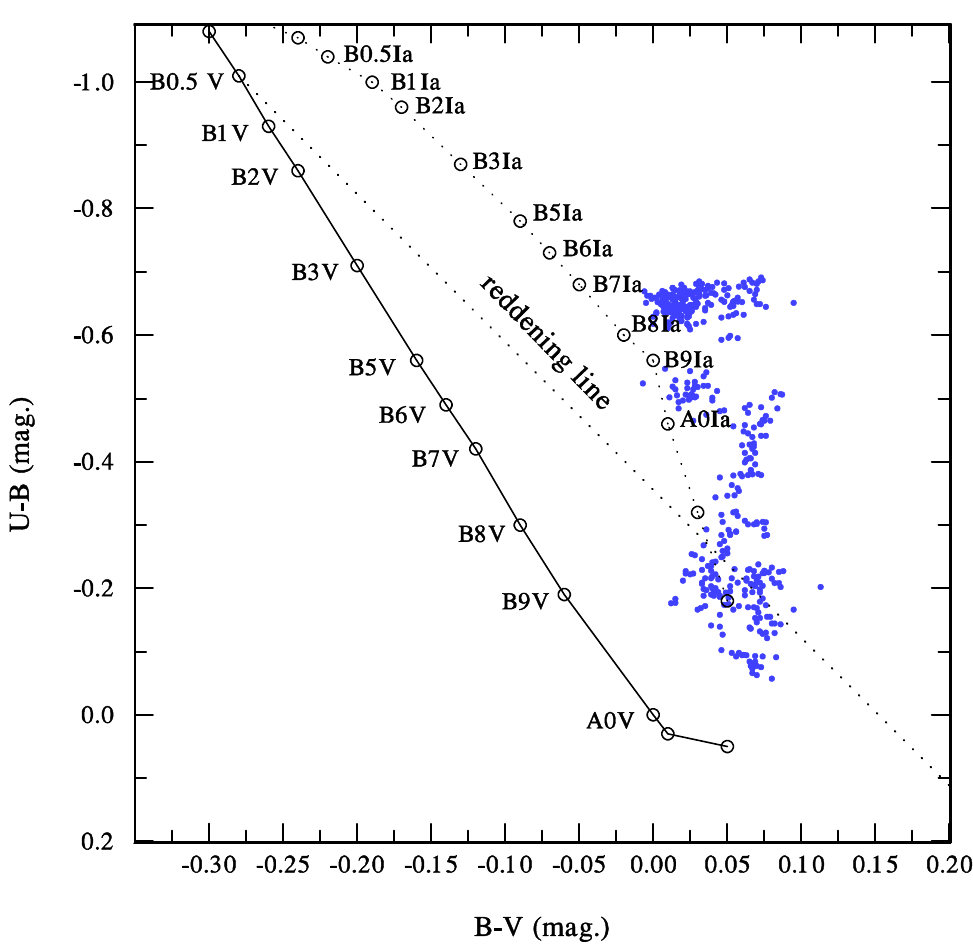}
\includegraphics[angle=0,scale=0.95]{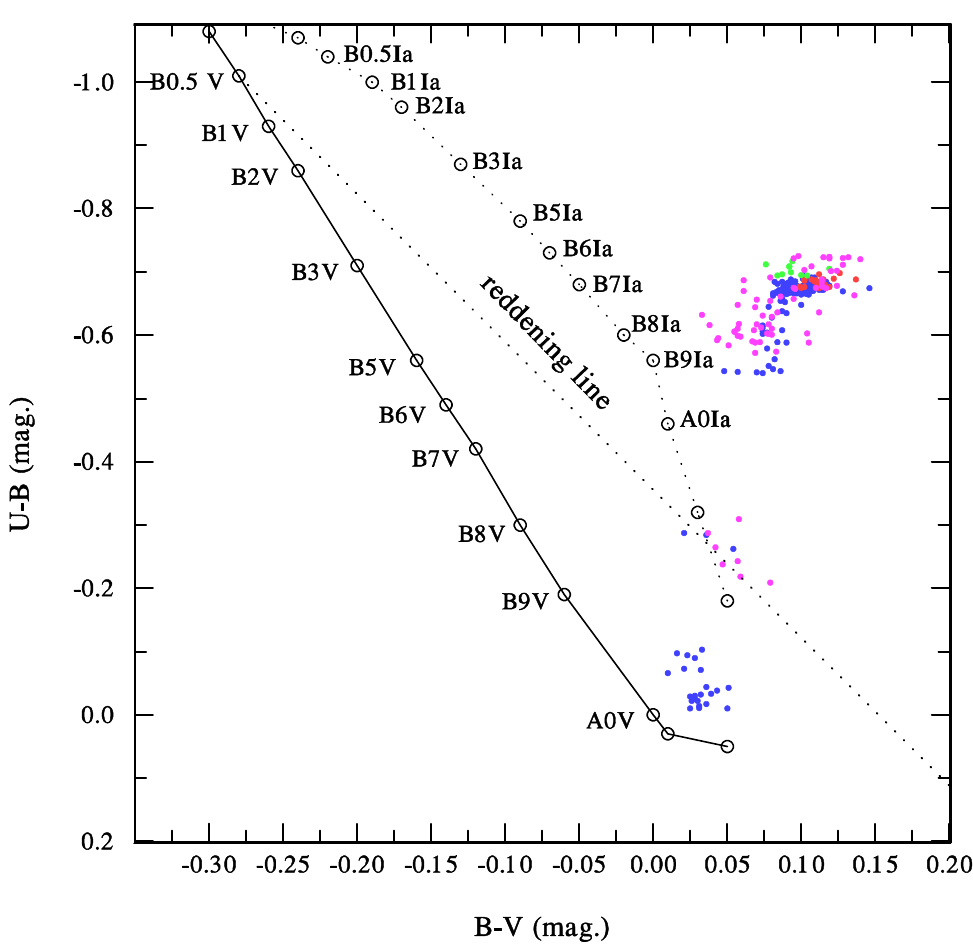}
\includegraphics[angle=0,scale=0.95]{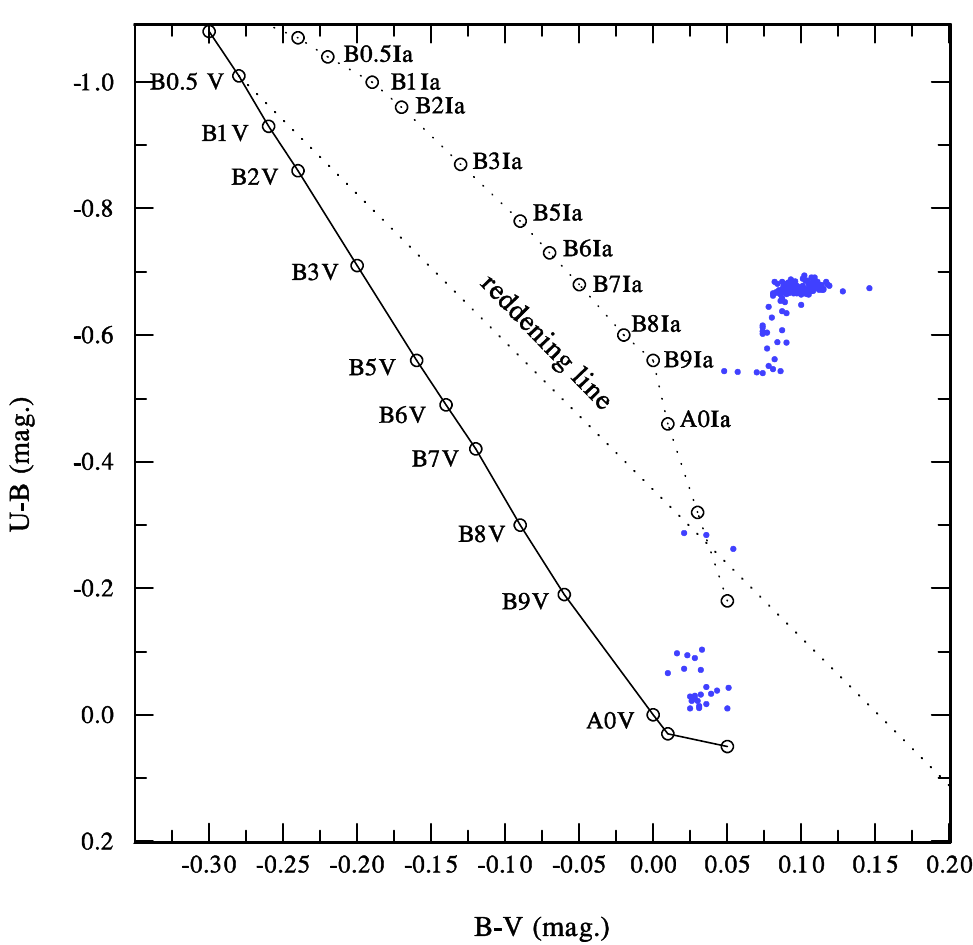}
\caption{ \ub\ vs. \bv\ diagram for several distinct data subsets.
Top panels: Older data until JD 2450000 (left). All-sky observations
are shown as black circles, and data from stations defined
in Table~\ref{jouphot} are denoted as follows:
01 (blue), 04 (green), 12 (red), and 26 (magenta).
More recent data from the interval of secular
light brightening, all from station 01 (blue) (right).
The bottom panels show \ubv\ observations from
the two sharp increases in the emission-line strength accompanied
by light decreases. Time interval JD~2452741 -- 2453309, which
covers the first sharp light decrease (left, cf. Figs.~\ref{vis-v} and \ref{all-n}.)
Data from the time interval JD~2455357 -- 2456094 corresponding
to the second sharp increase in the emission-line strength (right).
Data from stations 1, 30, 66, and 89 of Table~\ref{jouphot} are shown
by blue, red, green, and magenta dots, respectively. The main sequence
and the supergiant sequence based on data from \citet{golay74} (pp. 79-80)
are shown, as is the reddening line.}
\label{ubbv}
\end{figure*}

 Many systematically studied Be stars are known to exhibit always the same
and a rather clear type of either a positive or an inverse correlation
between the long-term brightness variations, a~characteristic type of
behaviour in the colour-colour diagram, and the Balmer emission-line strength
as defined by \citet{hvar83,hec2000}. He identified these two types
of correlation as an aspect effect.
For Be stars with an~inverse type of correlation, light decreases are followed
by the rise of the Balmer emission-line strength and by a~shift along
the main sequence towards later spectral subclasses in the \ub\ versus \bv\
diagram. For a~positive type of correlation, the brightenings are followed
by the rise of the emission strength and a~shift from the main sequence
towards the supergiant sequence in the \ub\ versus \bv\ diagram. The inverse
correlation is observed for stars that are seen more or less equator-on (a growing
gaseous envelope is attenuating the light of the central object), while
the positive correlation is observed for stars that are seen more pole-on (inner
optically thick parts of the growing envelope mimic an~apparent increase in
the stellar radius). Several examples of both types of correlation in
the colour-colour diagram can be found, for instance, in Fig.~2 of
\citet{bozic2013}.

  The situation is dramatically different for \ve. The \ubv\ observations
accumulated over several decades cover a large part of the whole
colour-colour diagram, with a single clear pattern.

  To understand better what is going on, we investigated
the colour-colour diagrams for different segments of the long-term
changes. Figure~\ref{ubbv} shows the colour changes separately for
the old data secured before JD~2450000, for more recent data from the secular
brightness increase (observations after JD~2457000), and for
observations covering two episodes of a rapid increase and decrease in the \ha emission
associated with sharp light decreases. The
pattern is remarkably similar for both these episodes. Formally, it looks like
a~positive correlation. However, the phases of minimum brightness and
maximum strength of the emission correspond to data that are close
to the main sequence, even below it when the reddening is considered.
The older data are all clustered above the supergiant sequence, while
the recent data lie along the supergiant sequence for late-B and early-A
spectral classes.
All this indicates that we observe a combination of several different types
of long-term changes.

\section{Duplicity of \ve}

\setcounter{table}{5}
\begin{table}
\begin{flushleft}
\caption{Orbital solutions based on the \ha emission RVs.}
\label{sol}
\begin{tabular}{lccrrccc}
\hline\hline\noalign{\smallskip}
 Element                   &All RVs        &Hi-res. spectra only\\
\noalign{\smallskip}\hline\noalign{\smallskip}
$P$ (d)                    &$192.91\pm0.18$& 192.91 fixed       \\
$T_{\rm super.conj.} ^*$   &$56318.5\pm2.2$&$56316.2\pm2.9$     \\
 $e$                       &0.0 fixed      & 0.0 fixed          \\
$\gamma$ (\ks)             & $-6.27\pm0.31$& $-5.52\pm0.44$     \\
$K_1$    (\ks)             & $6.33\pm0.41$ & $6.26\pm0.61$      \\
No. of RVs                 & 172           & 38                 \\
rms (\ks)                  & 3.92          &  2.63              \\
\hline\noalign{\smallskip}
\end{tabular}
\end{flushleft}
\tablefoot{$^*$) All epochs are in HJD-2400000.}
\end{table}

\begin{figure}
\resizebox{\hsize}{!}{\includegraphics{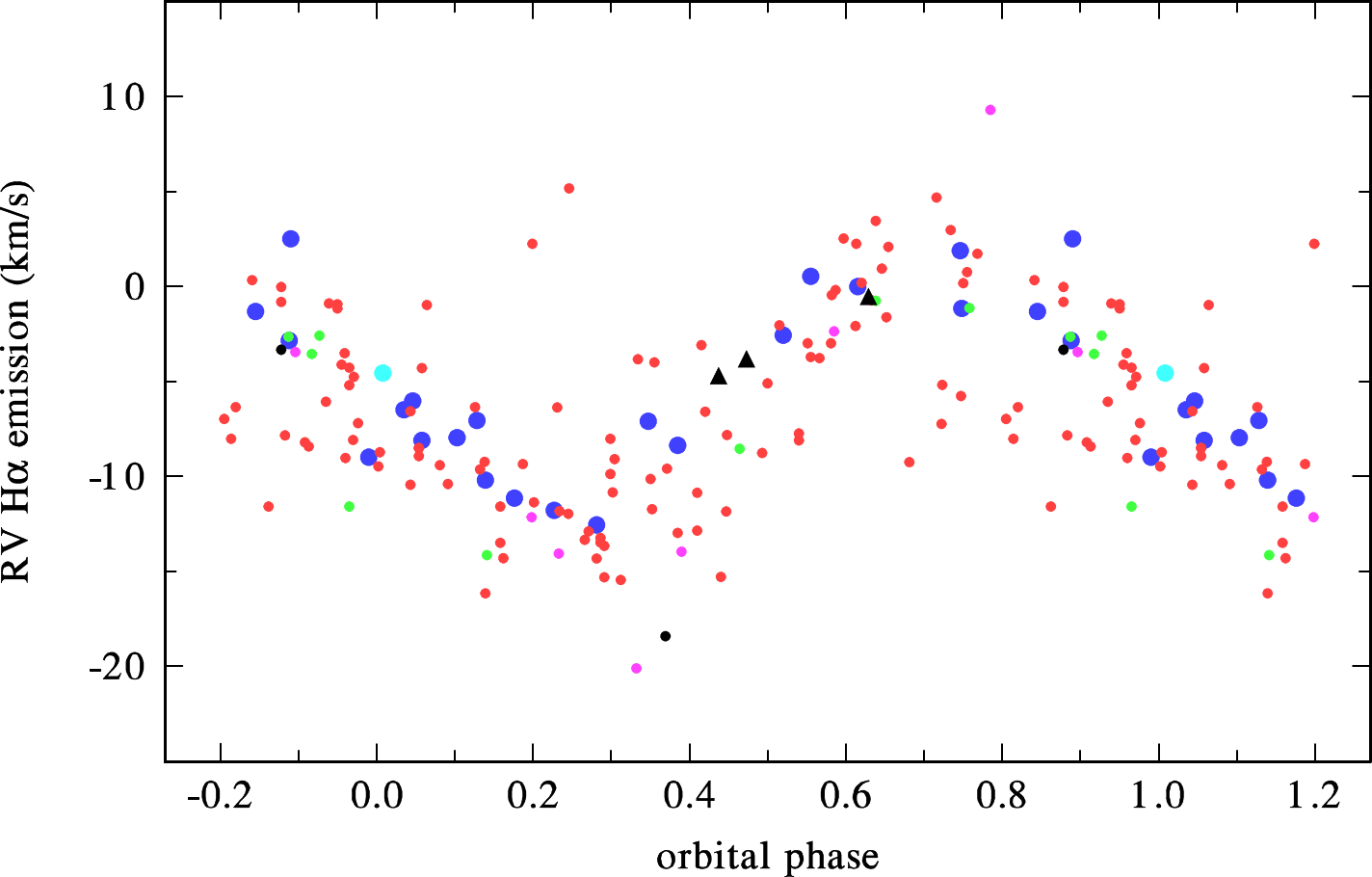}}
\resizebox{\hsize}{!}{\includegraphics{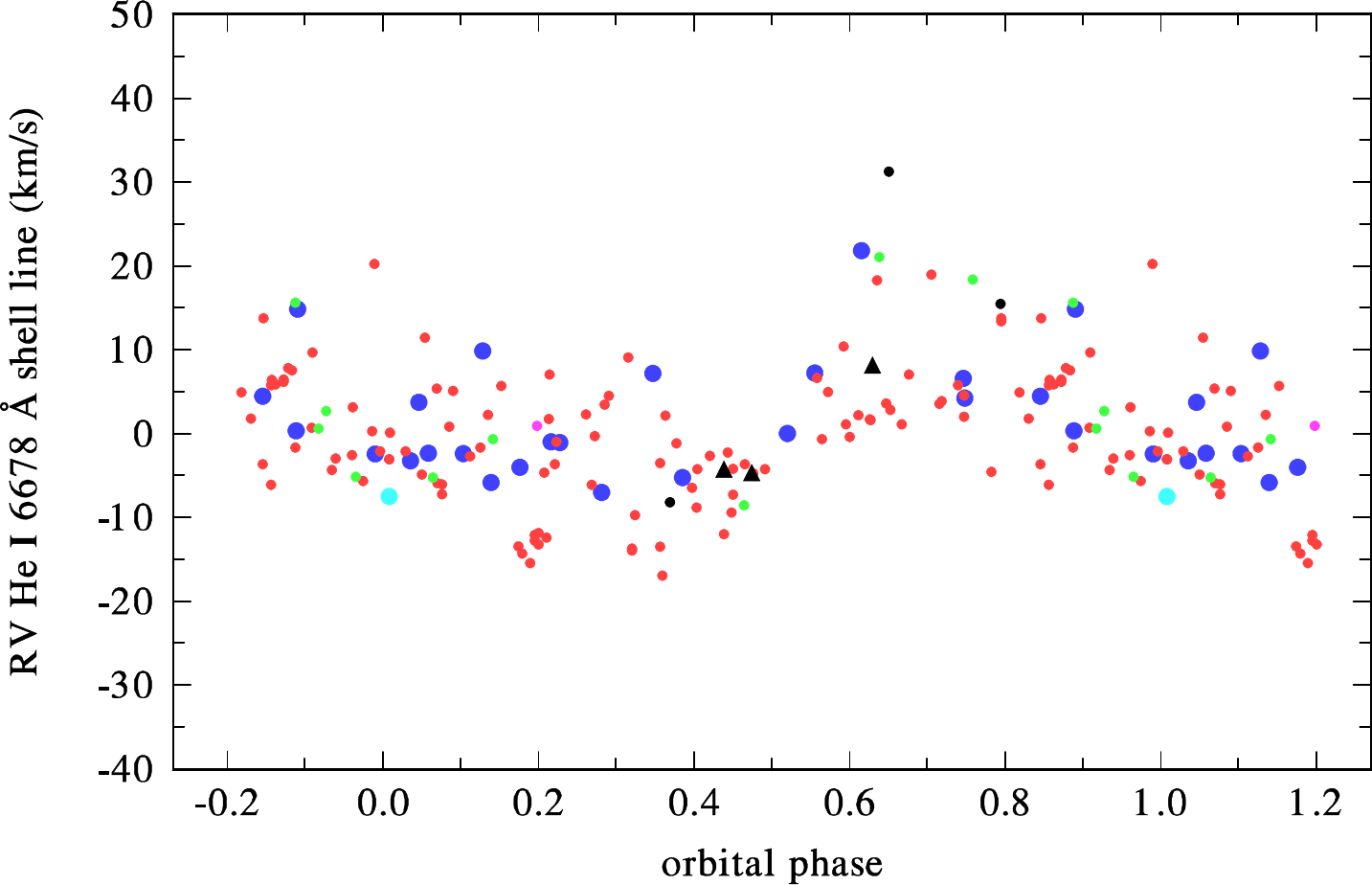}}
\caption{Radial-velocity curve corresponding to the orbital solution, based on
RVs measured on the steep wings of the \ha emission, plotted for
phases from ephemeris (top; \ref{efem}).
Orbital curve based on RV of the \ion{He}{i}~6678~\AA\ shell line
prewhitened for the long-term changes, plotted for the same ephemeris (bottom).
Data from individual instruments are shown by different symbols. The circles 
are the same as in Fig. 2, and black triangle shows CTIO.}
\label{orbit}
\end{figure}

The idea that duplicity can be an important factor for the very existence
of the Be phenomenon is not new. \citet{plahor69,kriz69} and
\citet{plavec70} have argued that at least some Be stars could be binaries
that are observed in the later phases of mass exchange between the binary components.
\citet{krizhec75} and \citet{hk76} formulated the general hypothesis that
Be stars are mass-accreting components of binaries and showed that this idea 
can also explain several types of time variations observed for Be
stars.  Additional arguments were provided by \citet{plavec76a} and
\citet{peters76}. However, as pointed out already by \citet{plavec76b},
if all Be stars have Roche-lobe filling secondaries,
more eclipsing binaries should be observed among them. Later investigations also led to the
finding that the presence of Roche-lobe filling secondaries can be excluded
for some Be stars that were found to be spectroscopic binaries, such as V744~Her = 88~Her
\citep{zarf7a,zarf7b} or V439~Her = 4~Her \citep{zarf5,zarf6}.
This led \citet{pols91} to suggest that many
Be stars might be objects created by large-scale mass transfer that were
observed in phases after the mass transfer ceased. The expected
secondaries of such objects would be hot compact stars, white dwarfs
in some cases. These are the most easily detectable sources in far-UV spectra. Evidence
for a hot secondary to the well-known Be binary $\varphi$~Per was found
first from the antiphase variation in the \ion{He}{ii}~4686~\AA\ emission
seen in the photographic spectra \citep{poeckert81} and later from
\ion{He}{i}~6678~\AA\ emission in the electronic spectra obtained
by \citet{gies93}. Its ultimate direct detection as an O~VI subdwarf
came from the study of the far-UV spectra from the Hubble Space Telescope by
\citet{gies98}. The secondary was then resolved with optical spectro-interferometry
by \citet{mourard2015}. Detections for several other systems followed.
\citet{wang2018} carried out a systematic search for the presence of hot
secondaries and summarised our knowledge of already known cases. \citet{wang2021}
detected nine new Be+sdO binaries from analyses of the Hubble Space Observatory
spectra, and \citet{klement2022} reported the first interferometric detection and
signatures of the orbital motion for three known Be+sdO systems. On the other
hand, \citet{boden2020} carried out a systematic search for Be stars with
main-sequence secondaries, with a completely null result. This constitutes
indirect evidence that the mass exchange is or was behind the formation
of binaries with Be primaries. \citet{hast2021} carried out evolutionary
calculations of mass exchange in binaries in an effort to set some limits
on the fraction of Be stars produced by binary interaction. They found
that under certain conditions, this fraction can be quite high.

\begin{figure}
\resizebox{\hsize}{!}{\includegraphics{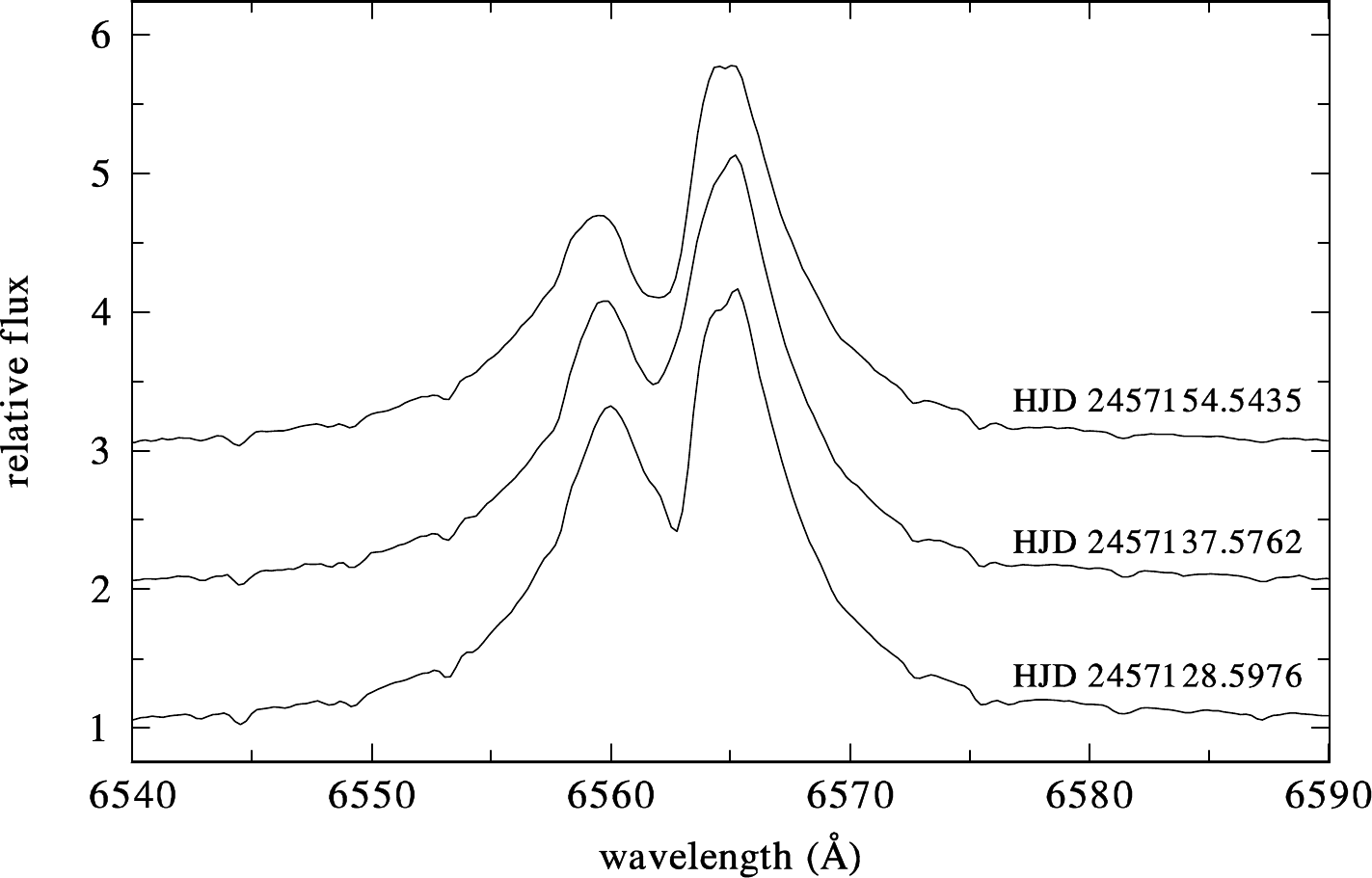}}
\caption{Three Ond\v{r}ejov \ha profiles. They are mutually shifted in
ordinate by 1.0 of the continuum level for better clarity.
Profiles from HJD~2457128.5976 and 2457137.5762 have anomalously positive
RVs of the emission wings, which stem from the episode of a large strengthening
of the emission. The next profile, from HJD~2457154.5435, has a~normal
orbital RV.}
\label{haprofs}
\end{figure}

\begin{table}
\begin{flushleft}
\caption{Possible properties of the binary system:
Mass of the secondary $M_2$, mass ratio $M_2/M_1$, semi-amplitude of
the RV curve of the secondary $K_2$, and the semi-major axis $a$ for
several possible orbital inclinations $i$. The mass of the primary was
assumed to be $M_1=16.9$~\Mnom\ after \citet{zorec2016}.}
\label{secmass}
\begin{tabular}{lccrrccc}
\hline\hline\noalign{\smallskip}
  $i$     & $ M_2$ &$M_2/M_1$& $K_2$  & $a$\\
($^\circ$)& (\Mnom)   &      & (\ks)  & (\Rnom) \\
\noalign{\smallskip}\hline\noalign{\smallskip}
 90.0  & 1.171 &  0.0693 & 90.43  &    368.69   \\
 85.0  & 1.175 &  0.0695 & 90.07  &    368.72   \\
 80.0  & 1.189 &  0.0704 & 88.99  &    368.82   \\
 70.0  & 1.249 &  0.0739 & 84.73  &    369.23   \\
 60.0  & 1.361 &  0.0805 & 77.77  &    369.98   \\
\hline\noalign{\smallskip}
\end{tabular}
\end{flushleft}
\tablefoot{We express the values of masses and radii
in the nominal values \Mnom, and \Rnom\ as defined by \citet{prsa2016}.}
\end{table}

One always has to be cautious when analysing binaries with clear signatures
of the presence of circumstellar matter in the system. The experience from
our previous studies of individual Be stars \citep{bozic95,zarf18,zarf20,
zarf21,zarf24,zarf26} shows that the binary nature of particular Be stars
is most easily detected via periodic RV variations of the steep
emission wings of the \ha line and often also via the periodic changes
in the $V/R$ ratio of the double Balmer emission lines.

Period analyses of all \ha emission-line RVs of \ve, using
both the \citet{deeming75} and \citet{stelling78} methods, revealed that
the RV of the \ha\ emission wings varies with a period of 193~d
and a~semi-amplitude of $\sim 5$ \ks.
The same periodicity is also detected in the RV of the \ha absorption core
and in the absorption RVs of \ion{Si}{ii} doublet at 6347 and 6371~\AA\,
and \ion{Fe}{ii}~6456~\AA\  after long-term changes are removed.

Using the program \fotel \citep{fotel1,fotel2}, we derived the circular-orbit
orbital elements for all \ha emission-wing RVs and for those from the
high-resolution spectra alone. The mutual agreement of the two solutions is
very satisfactory. They are presented in Table~\ref{sol}, and
the corresponding RV curve is plotted in Fig.~\ref{orbit}.
In the rest of this study, we adopt the following linear ephemeris:

\begin{equation}
T_{\rm super.conj.}={\rm HJD}\,2456318.5+192\fd91\times E \label{efem}
\end{equation}

\noindent based on the solution for all spectra.

 To be fair, we note that some deviations from the mean RV curve in the upper
panel of Fig.~\ref{orbit} are rather large. This is, for instance,
the case of two Ond\v{r}ejov spectra taken on HJD~2457128.6 and 2457137.6,
when a very steep rise of the emission strength
had occurred. We remeasured these spectra several times, but the result was the
same. Their RVs are almost in anti-phase to the orbital RV curve near phase 0.3.
We show the corresponding line profiles in Fig.~\ref{haprofs} together with
another profile, taken about two weeks later, which already gives a~RV in accord
with the orbital motion. The two peculiar RVs were given zero weight in the
orbital solution.

We tentatively adopted the mass of the Be primary after \citet{zorec2016} and
estimated the basic properties of the system for several
possible orbital inclinations. Because no lines of the secondary were detected
in the optical spectra and because no companions to Be stars were ever found
among main-sequence objects \citep{boden2020}, we conclude that the secondary
is not a Roche-lobe filling object, but most probably a hot subdwarf star or
white dwarf. It should be looked for in the far-UV spectral region.
For the Gaia DR2 parallax of 0\farcs0007059, the projected angular separation
of the binary components is 0\farcs0048, which might be resolved with 
present-day optical interferometers such as the currently tested interferometer
SPICA \citep{spica2020}.

\section{Correlations between orbital and long-term changes}

\begin{figure}
\resizebox{\hsize}{!}{\includegraphics{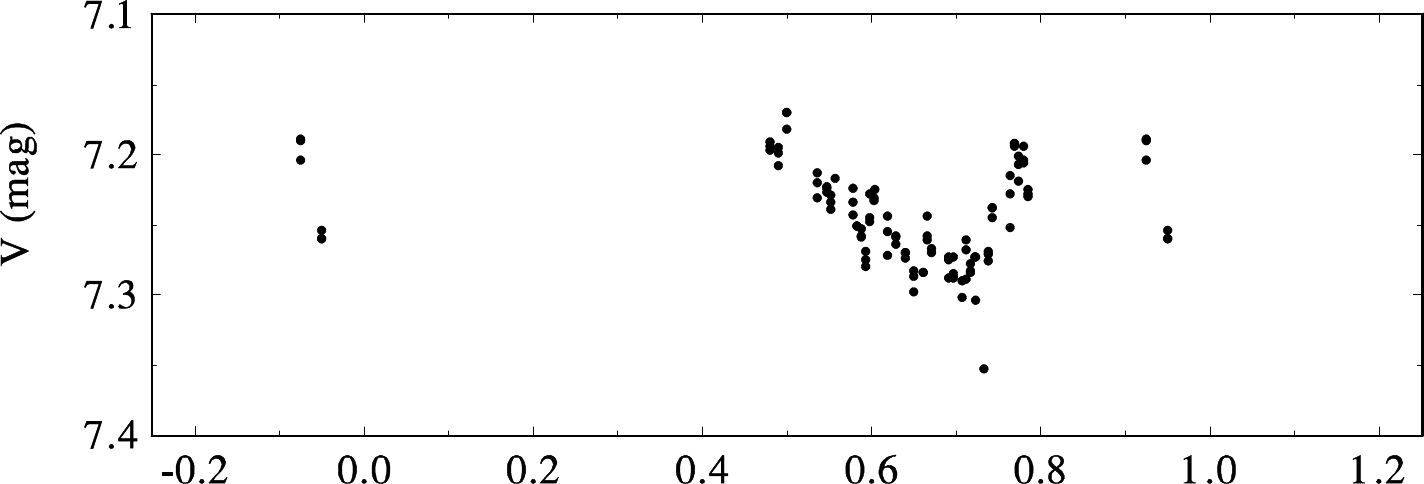}}
\resizebox{\hsize}{!}{\includegraphics{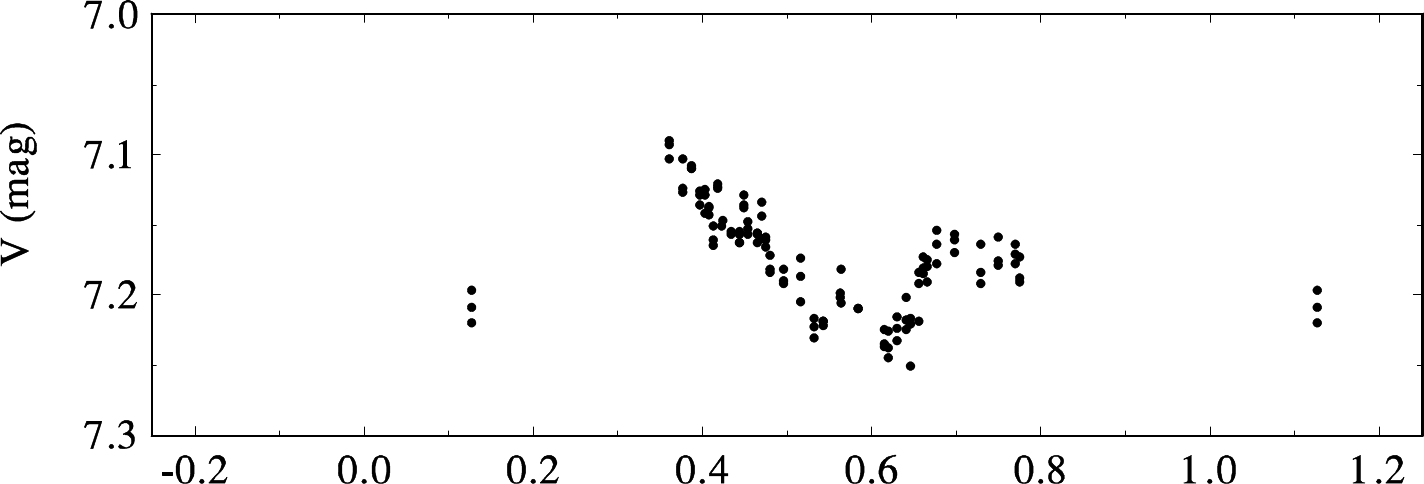}}
\resizebox{\hsize}{!}{\includegraphics{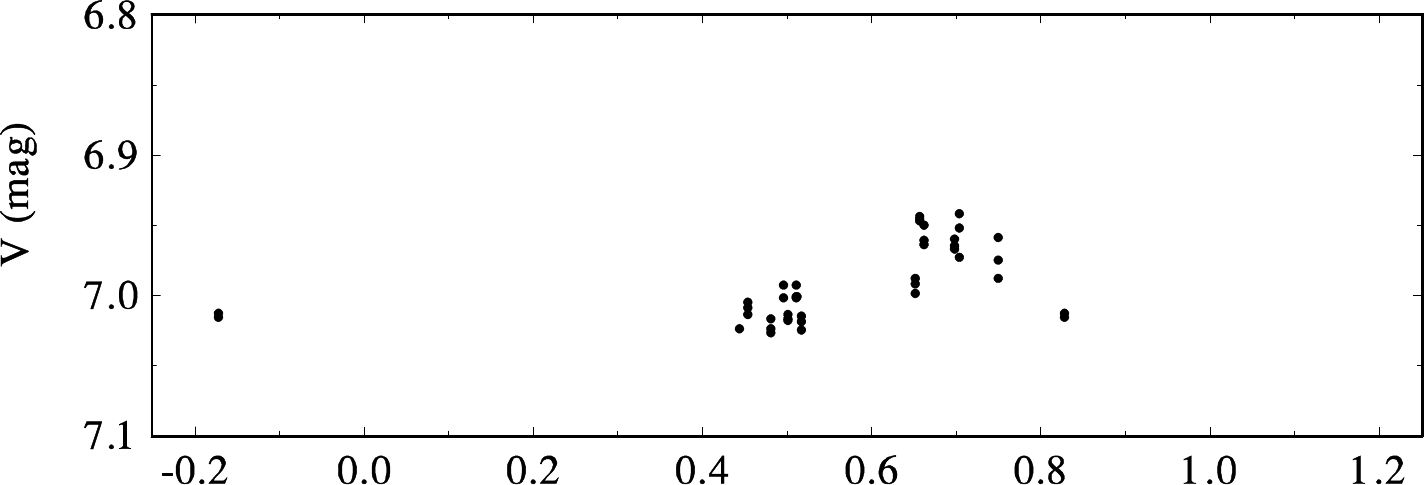}}
\resizebox{\hsize}{!}{\includegraphics{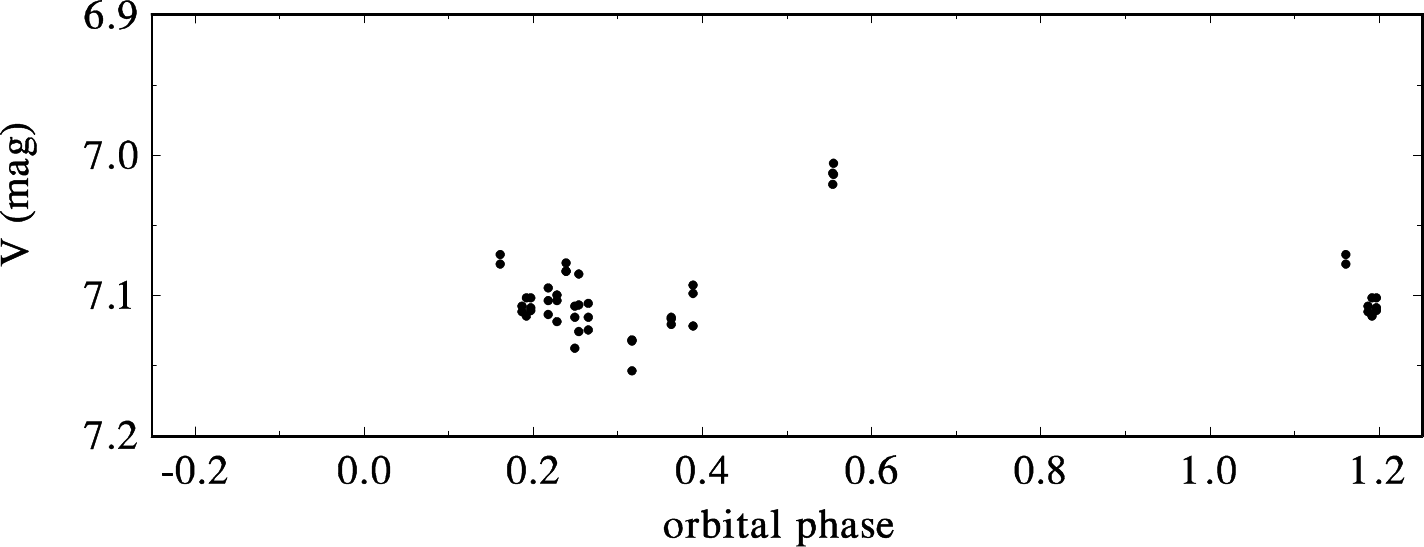}}
\caption{Phase plots of $V$ magnitude for several seasons of Hvar observations
for binary ephemeris (\ref{efem}). From top to bottom: Data from
JD~$2457568-57657$,
JD~$2457931-58079$,
JD~$2458318-58392$, and
JD~$2458664-58740$.}
\label{vorb}
\end{figure}

\begin{figure}
\resizebox{\hsize}{!}{\includegraphics{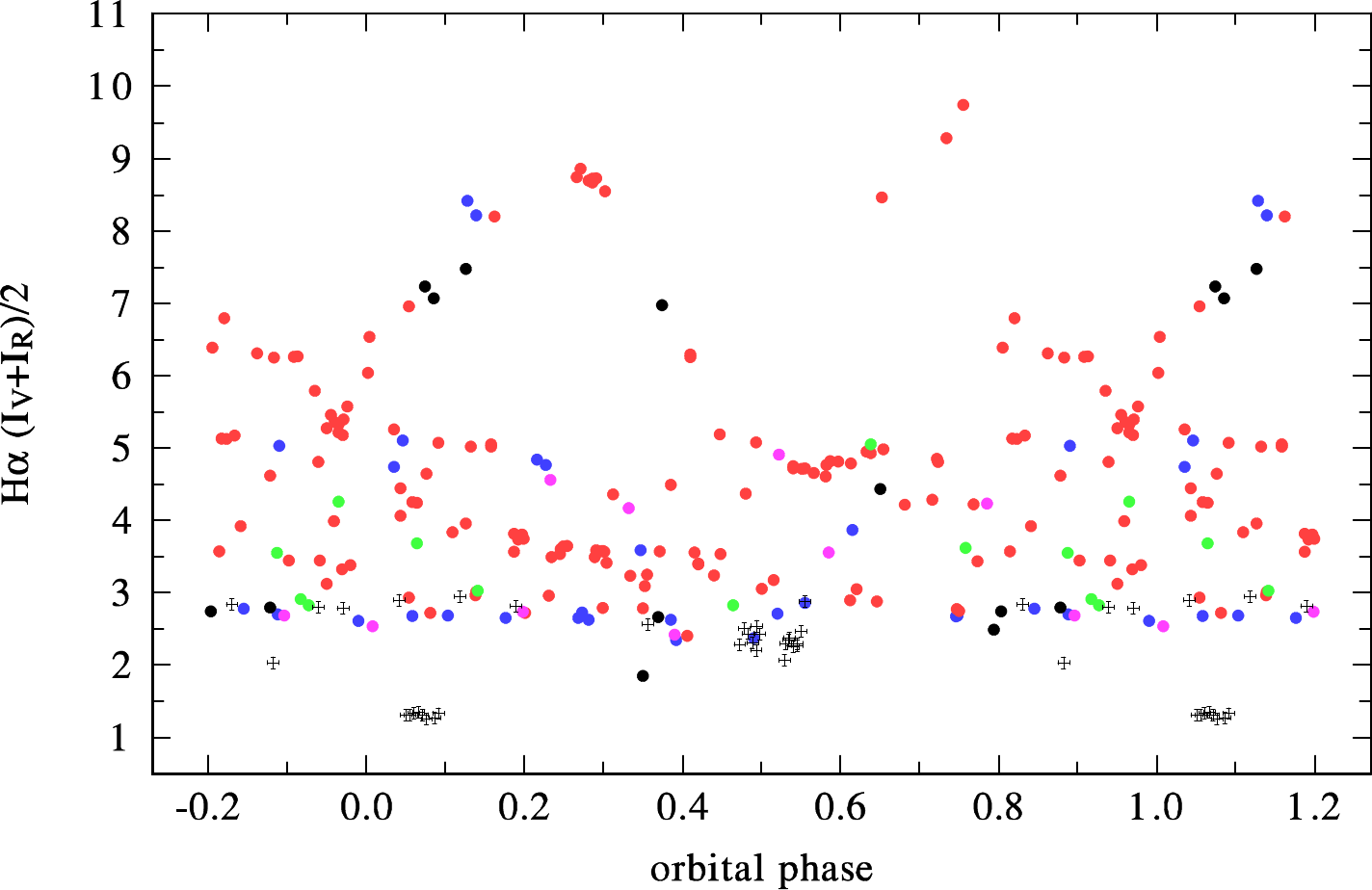}}
\caption{Phase plots of the strength of the \ha emission
for binary ephemeris (\ref{efem}). Data from individual sources are
denoted as follows: Circles in blue show DAO spectra, red circles show OND spectra, green circles represent BESO spectra,
magenta circles show BeSS, and black circles show Castanet spectra. The black crosses show data from the literature.}
\label{haemisph}
\end{figure}

Inspection of the light, colour, and emission-line strength variations seems to
indicate that the rapid episodes of large changes such as those near JD~2452900
or JD~2457300 occurred during one binary orbital period. To investigate the problem,
we plot phase diagrams of the variability of the $V$ magnitude for several
more recent shorter time intervals in Fig.~\ref{vorb}. The two
large light decreases that are accompanied by strong increases in the \ha emission-line
strength apparently occurred around phases of elongations, with the Be primary receding from us.
At the same time, the plot shows that in another observing season, brightenings
were observed around similar orbital phases. The same is also confirmed by
a phase plot of the \ha emission-line strength; see Fig.~\ref{haemisph}.

\begin{figure}
\resizebox{\hsize}{!}{\includegraphics{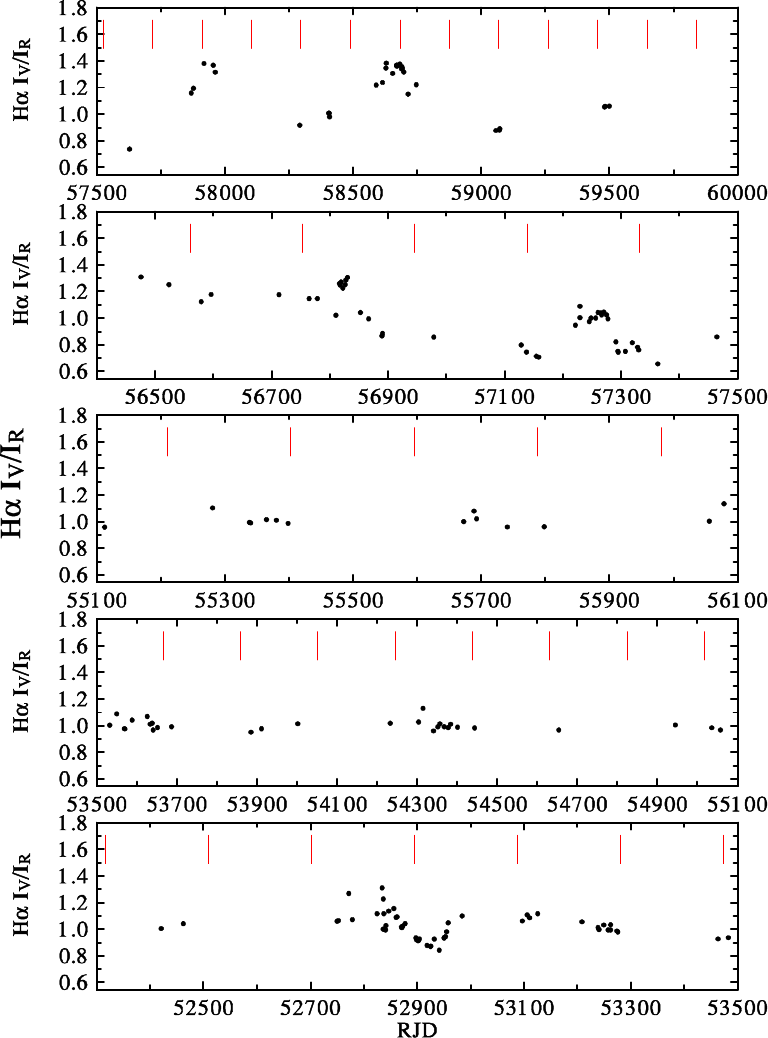}}
\caption{Enlarged subsets of time variability of the \ha $V/R$ ratio
with the instants of the expected phase-locked maxima predicted by the
orbital ephemeris (\ref{efem}).}
\label{vrsets}
\end{figure}

  We also investigated the time behaviour of the $V/R$ ratio of the double
\ha emission. In this case as well, \va appears to be quite unusual.
As Figs.~\ref{all} and \ref{all-n} show, the $V/R$ variations are different
in different time intervals and do not recall either the long-term cyclic
changes known for Be stars with one-armed global oscillations or phase-locked
changes. In several panels of Fig.~\ref{vrsets} we show enlarged plots of
the $V/R$ changes. Instants of expected phase-locked $V/R$ maxima predicted
by the orbital ephemeris (\ref{efem}) are shown by vertical lines.

\section{Rapid changes}
\begin{figure}
\resizebox{\hsize}{!}{\includegraphics[angle=0]{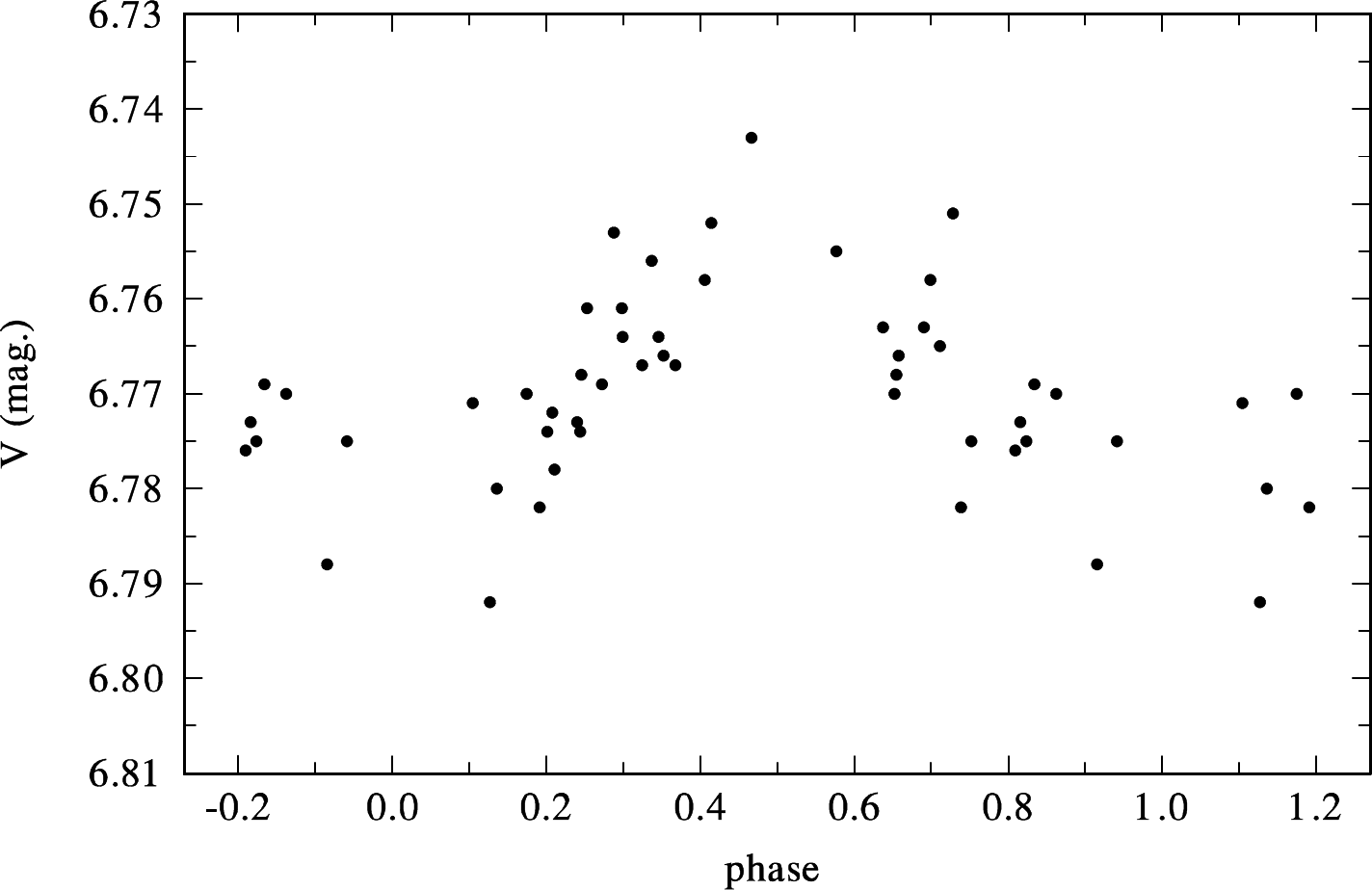}}
\caption{Possibly periodic rapid light changes based on \citet{lynds59}
$V$ magnitude photometry and plotted for ephemeris (\ref{efelynds}).}
\label{rapid}
\end{figure}

Although we collected a large number of photometric observations of \ve,
their time distribution is not suitable for a search for rapid periodic changes. Perhaps the only observations suitable for
such a search are the early $V$ observations by \citet{lynds59}.
He himself concluded that his observations definitively indicate
variations in brightness, which appear to be somewhat erratic, however, and
no period could be found. The observations were secured within one month
during a~time interval that was  not affected by secular variations. Our period analysis
revealed sinusoidal variations with a semi-amplitude of 0\m0139(29).
A~least-squares fit led to ephemeris (\ref{efelynds}), the rms of one
observation being 0\m0067. The corresponding phase plot is shown in Fig.~\ref{rapid}.

\begin{equation}
T_{\rm light\, min.}={\rm HJD}\,2436450.8053(92)+0\fd64827(50)\times E. \label{efelynds}
\end{equation}

This indicates that a scatter of at least 0\m03 is to be expected in individual
observations on longer timescales.

\citet{lefe2009} carried out an~automatic period search in the Hipparcos \hp\
photometry to find new periodic variables among OB stars. They identified
\va as a~possible slowly pulsating B star (SPB) with a period of 7\fd752.
We cannot confirm their result. They apparently did not take the secular
light change in \hp\ photometry into account; see the upper panel of Fig.~\ref{all} here.

\citet{zorec2016} estimated the following physical properties of the Be component:\\
\tef = ($30120\pm2540$)~K, \lgg = ($4.08\pm0.40$) [cgs],
mass M = ($16.9\pm2.7$)~\ms, \vsin = ($207\pm18$)~\ks,
critical rotational velocity $v=(517\pm64)$~\ks, and the
inclination of the rotational axis $i=(37^\circ\pm9^\circ)$.
For these values, the period 0\fd648 appears as a reasonable rotation period of the Be component.
In passing we note that at the time of writing, the star has not been observed
by the TESS satellite.

\section{Fourth timescale}
\citet{hec98} has called attention to the fact that the brightness of the
Be star $\omega$~CMa outside of the episodes of brightenings accompanied
by the growth of emission-line strength (typical of the positive correlation
discussed above) has been decreasing secularly. His observation was later
confirmed with more recent photometry \citep{ghore2018, ghore2021}. These
authors and also \citet{marr2021}, who studied another Be star, V2048~Oph = 66~Oph,
modelled the secular variability and the episodes of brightening and
increases in the Balmer emission-line strength with some success, estimating
the required viscosity values for individual episodes, and also discussing
some limitations of their effort. The yellow light curve of V2048~Oph
is shown in the upper panel of Fig.~1 of \citet{marr2021}. It shows a~secular
light decrease between 1980 and 2000, occasionally interrupted by brightenings
reminiscent of a positive correlation. However, the strength of
the \ha emission is near its maximum over the same time interval of about
20 years, and only then does it gradually decrease. No emission has been
observed since about 2010. However, as \citet{marr2021} pointed out,
the outer parts of the disk are still seen in the radio wavelengths.

\begin{figure}
\resizebox{\hsize}{!}{\includegraphics[angle=0]{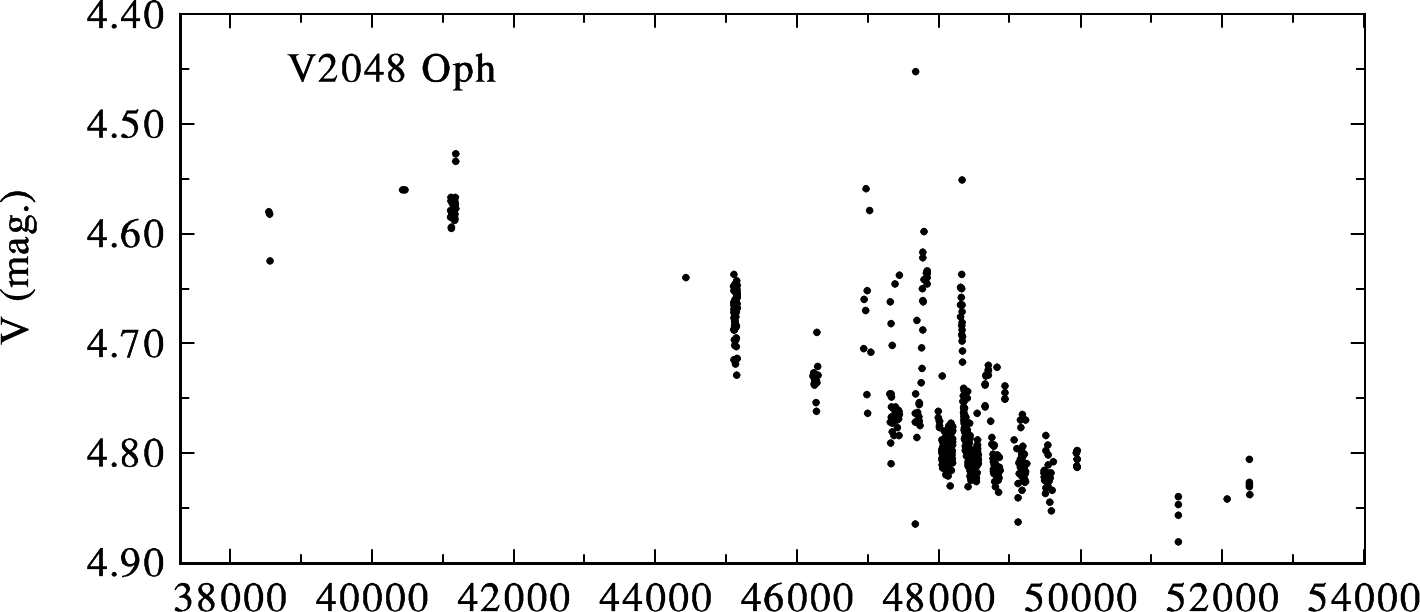}}
\resizebox{\hsize}{!}{\includegraphics[angle=0]{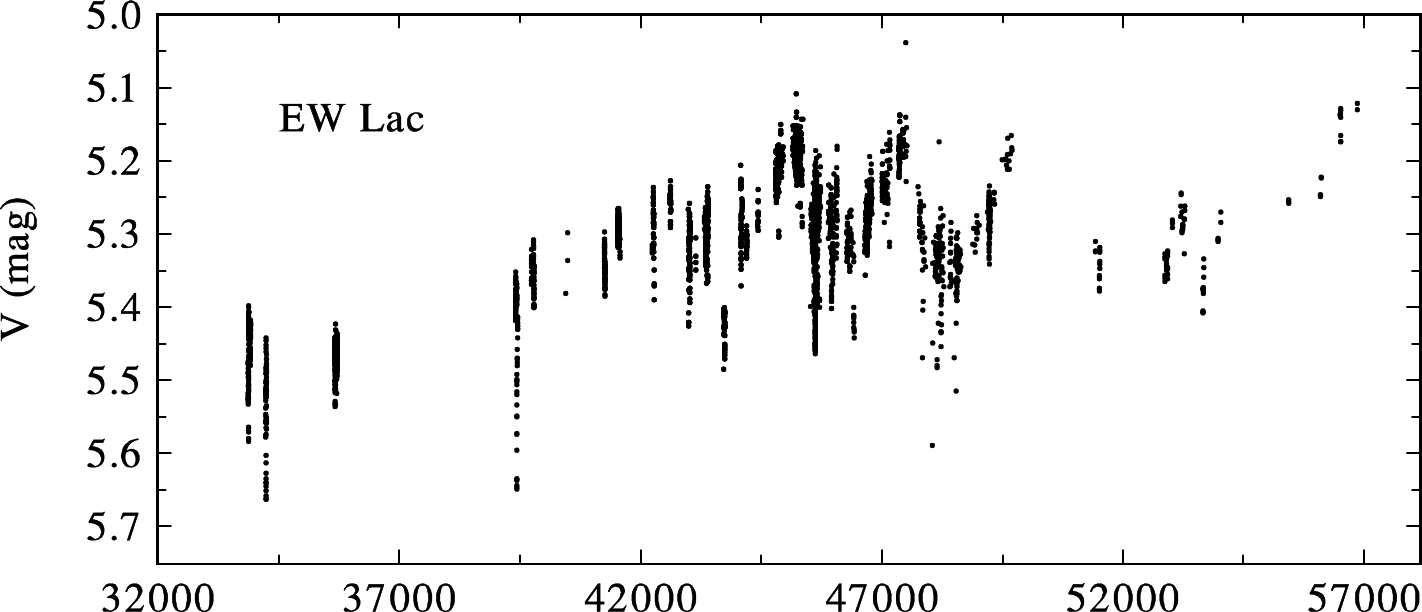}}
\resizebox{\hsize}{!}{\includegraphics[angle=0]{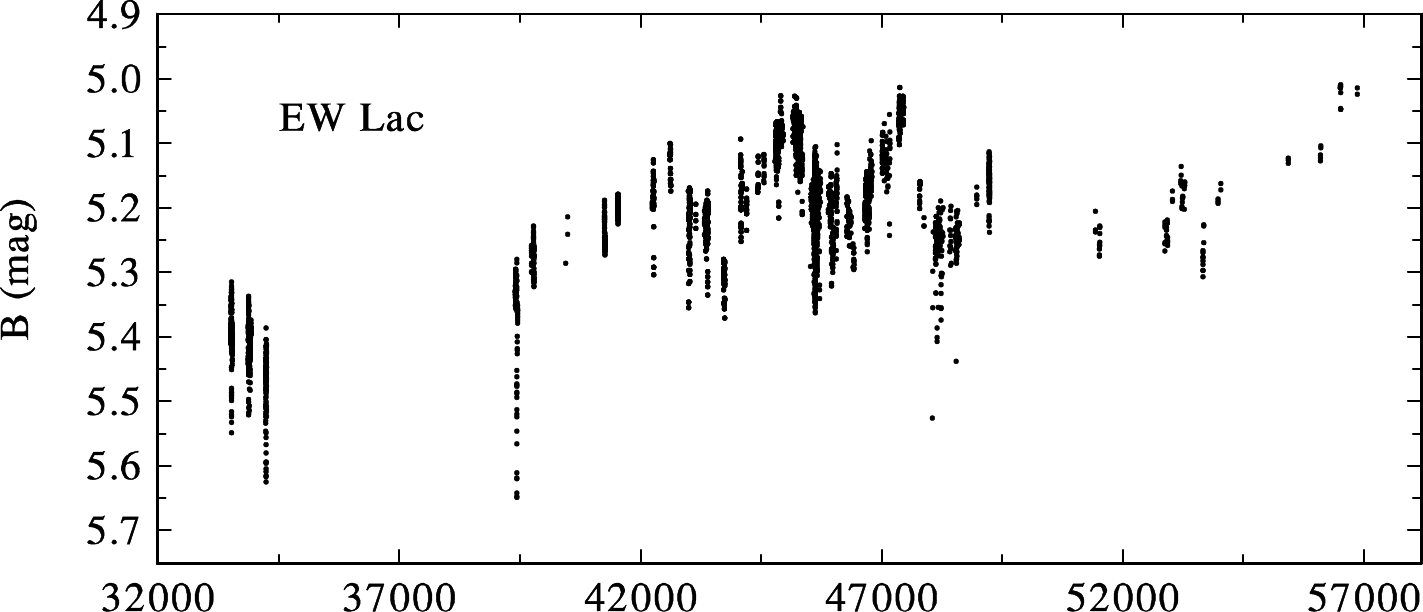}}
\resizebox{\hsize}{!}{\includegraphics{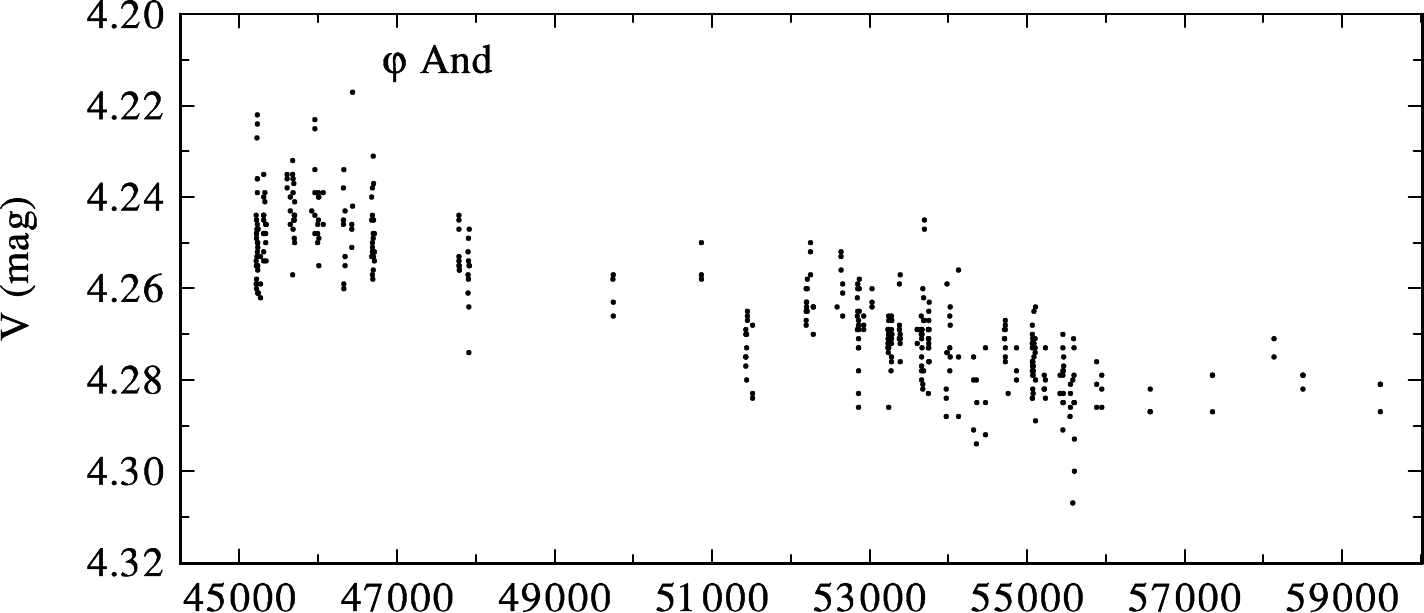}}
\resizebox{\hsize}{!}{\includegraphics{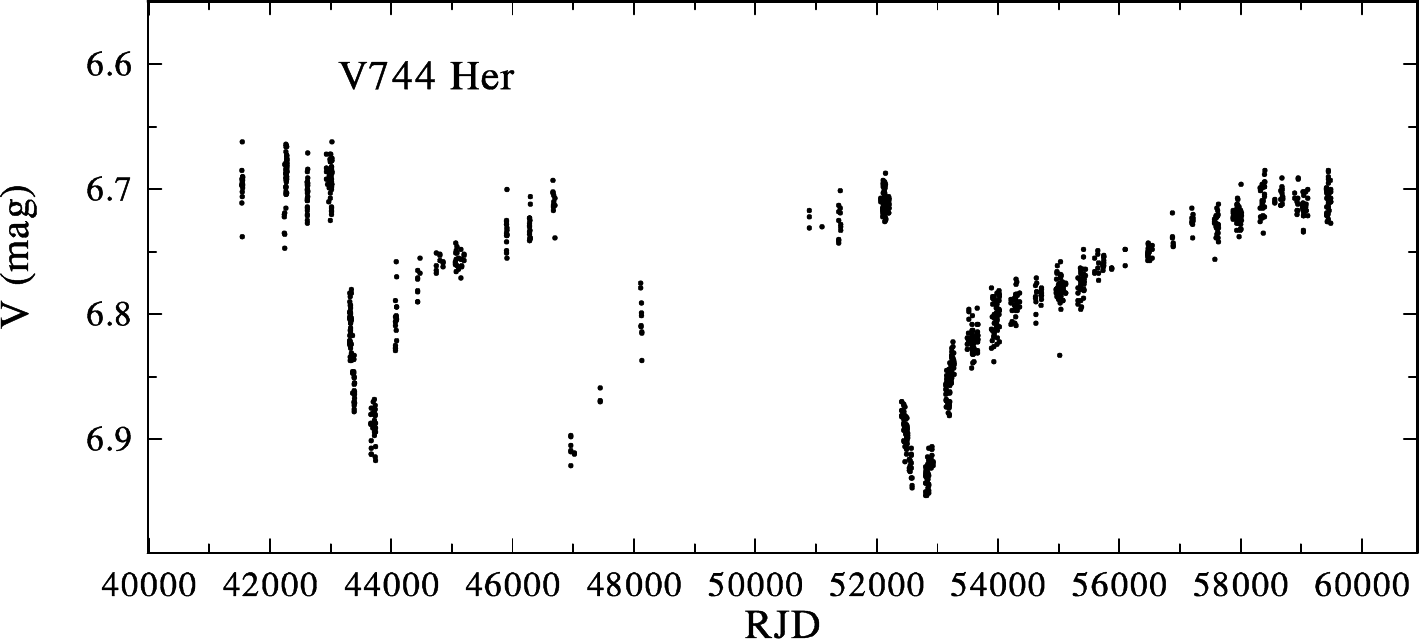}}
\caption{Secular photometric changes of several well-observed Be stars.}
\label{fourth}
\end{figure}

We collected and homogenised the $V$ photometry of V2048~Oph
from the archive of \ubv\ photometry provided by J.R.~Percy, from Hvar,
SPM, \citet{john66}, \citet{haupt74} and \citet{kozok85} and transformed the Hipparcos
\hp\ \citep{esa97} photometry into Johnson $V$ and the observations of
\cite{hill76} secured in the DAO photometric system into \ubv \ using the transformation
formul\ae\ provided by \citet{hecboz01}.  The $V$ light curve of V2048~Oph
based on the above-mentioned data sets is shown in the upper panel
of Fig.~\ref{fourth}.

A similar secular light decrease has also been reported for V744~Her = 88~Her,
a Be star with an inverse type of correlation \citep{hecboz2013}. We show
its light curve complemented by more recent observations, adapted from
Bo\v{z}i\'c et al. (in prep.) in Fig.~\ref{fourth} as well. In the same figure,
we also show the $V$ and $B$ light curves of the Be star EW~Lac = HD~217050
from another study in preparation. In this case, a secular increase
in brightness is observed. We note that the large scatter around the mean trend
is related to the known rapid light variability of EW~Lac on a timescale shorter than
one day. The fourth panel of Fig.~\ref{fourth} shows the plot of Hvar $V$ photometry
of $\varphi$~And, another Be star with a positive type of correlation.
A mild light decrease over several decades of observations is visible.

Searching the literature, we found a few more examples.
A secular light increase has also been observed for the Be star with a~positive
correlation $\gamma$~Cas \citep[][Fig.~5]{gcas2002} over nearly 30000~days.

Finally, as we have shown here, the secular light decrease of \va
has recently changed to a secular light increase.
All this shows the large variety of different possible evolutions
of the circumstellar disk, its replenishment, and a~gradual dissipation.

\section{Discussion}
In spite of the effort of several generations of stellar astronomers,
the engine leading to the occasional formation of circumstellar disks
around Be stars has not been firmly identified so far.
One  possible explanation is based on the idea that Be stars are
rapidly rotating non-radial pulsators (NRP) and that the additional
force needed to facilitate the outflow of gas and angular momentum transfer
from the stellar equator arises from a constructive interference of two
or more NRP modes
\citep{rivi98a,rivi98b,baade2017a,baade2017b,baade2020,borre2020,bartz2021}.
Especially the systematic photometries from space observatories were
used and analysed to support this conjecture. Confirmation of this
scenario would require the creation of new, self-consistent models,
however, which would show that Be stars are indeed pulsationally unstable over the whole
area that they occupy in the Hertzsprung-Russell (HR) diagram.
It should also be mentioned that \citet{baade2017b} warned that evidence
for constructive interference of pulsational modes for a larger number of Be stars
is lacking and pointed out additional problems such as the rotational splitting of modes
and the presence of rapid changes in the circumstellar envelopes during active phases.
An alternative view was suggested by \citet{hec98}, who argued that the dominant
period of rapid changes undergoes small cyclic changes. Modelling such
a situation, he found that a standard period analysis of a~corresponding series
of observations returns a multiperiodicity with several close periods.
Yet another possibility was suggested by \citet{bebin2002}, who argued that
the presence of a secondary can facilitate outflow from the equatorial parts
of the gaseous disk of the Be primary even in systems that are not filling
their Roche lobes. This idea is problematic, however, in that the effect is
rather small in the majority of cases.

 For a long time, various other suggestions have been made that
the Be phenomenon and observed variability patterns of Be stars might be causally
related to their binary nature
\citep[e.g. ][among others]{plahor69,krizhec75,hk76,bebin87,pols91,
  pano2018,boden2018,boden2020,langer2020,boden2021}. 
The approach of \citet{klement2019} is worth mentioning. These authors studied 
the spectral energy distribution of several Be stars and provided arguments that 
their disks had to be truncated by the Roche lobes. This constitutes 
an~indirect argument for the presence of companions to these objects.

  The phase-locked $V/R$ changes observed for several Be binaries represent another
interesting phenomenon. As already noted above, they are usually
observed roughly in phase with the orbital RV changes
of the Be stars in question \citep{zarf6,zarf7b,zarf16,zarf21,stefl2007}. A phase-locked
emission-line variation with a single maximum and minimum per orbital period
was also found by \citet{borre2020} for $\gamma$~Cas. These authors used
an interesting detection technique. They analysed local pixel-per-pixel line
fluxes across the \ha profile in a~series of higher-resolution BeSS spectra.
\citet{pano2018} modelled the phase-locked $V/R$ changes as global oscillations
in the circumstellar disks with two spiral patterns and concluded that the
phase-locked $V/R$ changes should exhibit two maxima and minima during one
orbital period. This clearly disagrees with the available observations
mentioned above. The only case for which a double-wave $V/R$ curve was detected is
V696~Mon = HR~2142 \citep{peters72,peters76}. There is a natural explanation
of roughly sinusoidal phase-locked $V/R$ changes in our view, as discussed
in Appendix~C of \citet{wolf2021}. The circumstellar disk probably
occupies almost the whole volume of the Roche lobe near
the orbital plane. This causes there to be more gaseous material in the part
of the disk facing the secondary than on the opposite side. Because the disk
rotates, more emission power is available on the side facing the secondary,
and this naturally leads to phase-locked $V/R$ changes that are in phase
with the RV changes of the Be component.

\begin{figure}
\resizebox{\hsize}{!}{\includegraphics[angle=0]{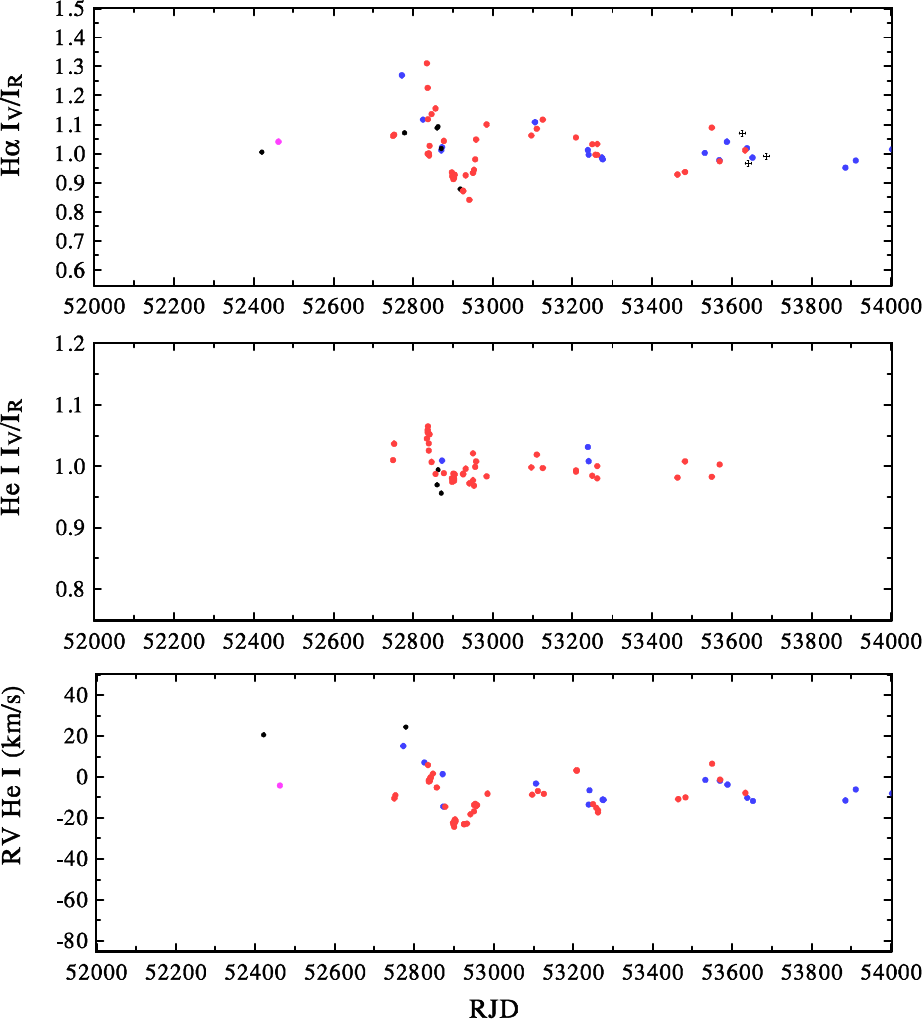}}
\resizebox{\hsize}{!}{\includegraphics[angle=0]{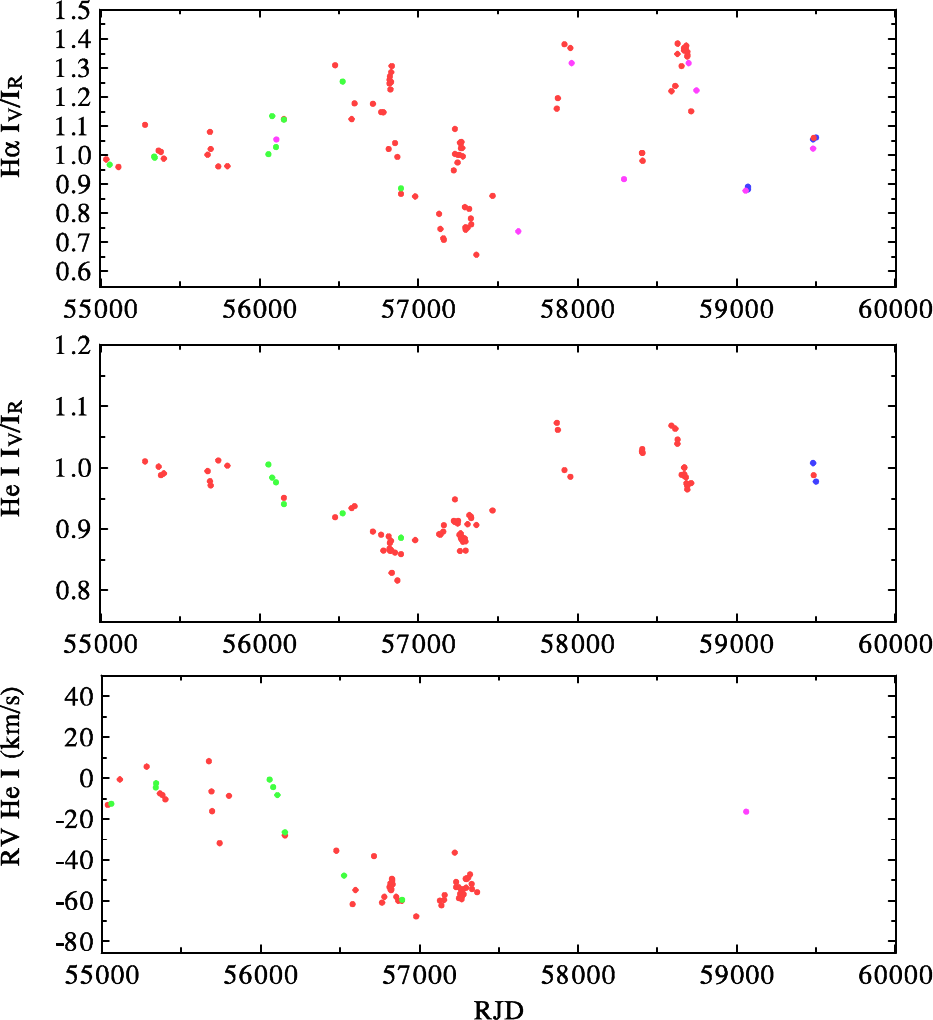}}
\caption{Apparent $V/R$ changes observed for the \ha and \he
emission lines intercompared for two time segments. The variation
in the shell He RV is also shown. An~apparently anti-phase behavior
is observed in the second time interval, when the shell RV becomes
quite negative and the He shell line blends with the $V$ peak of the
faint He emission. The same colours as in the previous time plots
are used to distinguish spectra from individual observatories.}
\label{fales}
\end{figure}

\begin{figure}
\resizebox{\hsize}{!}{\includegraphics[angle=0]{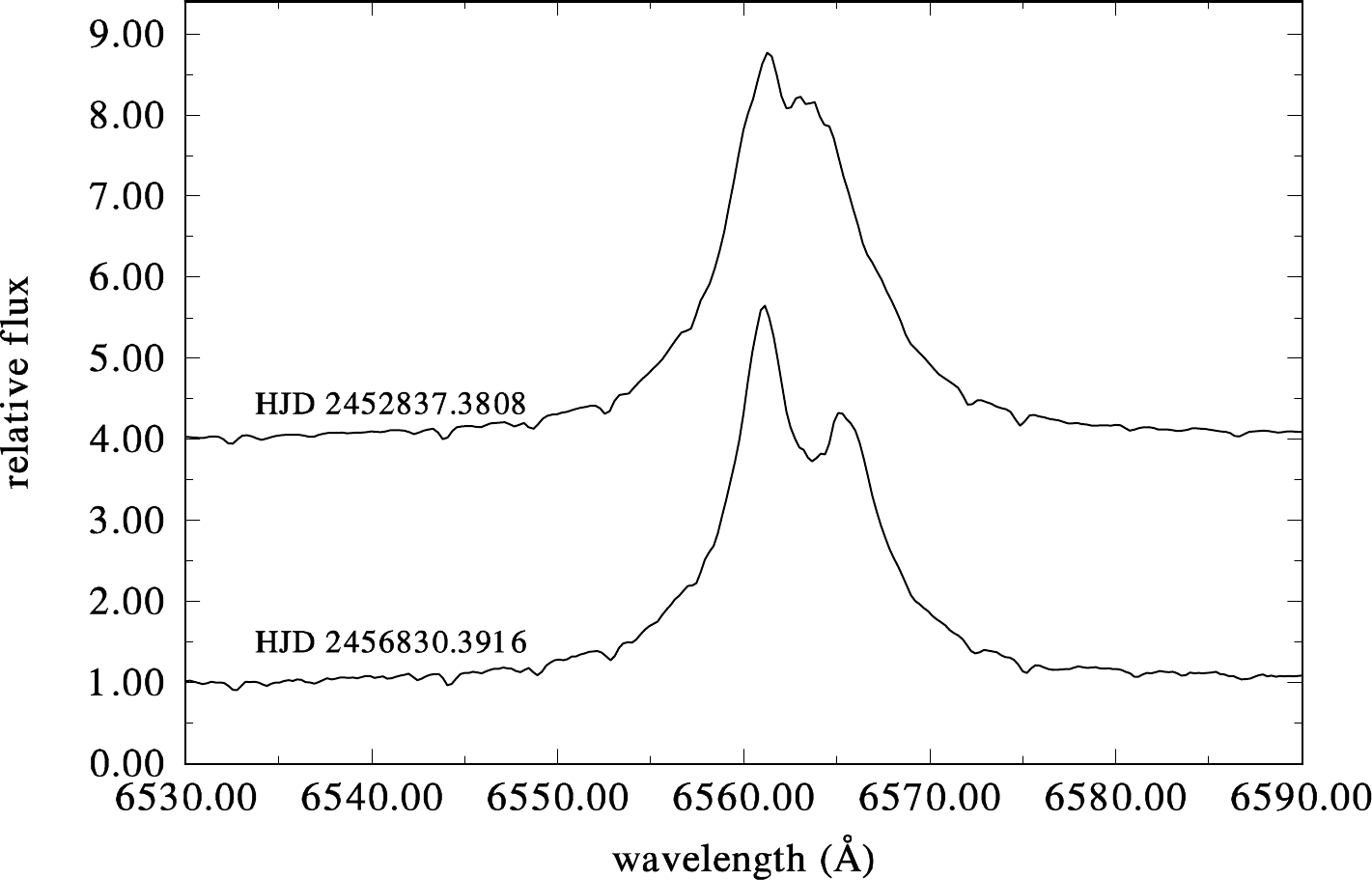}}
\resizebox{\hsize}{!}{\includegraphics[angle=0]{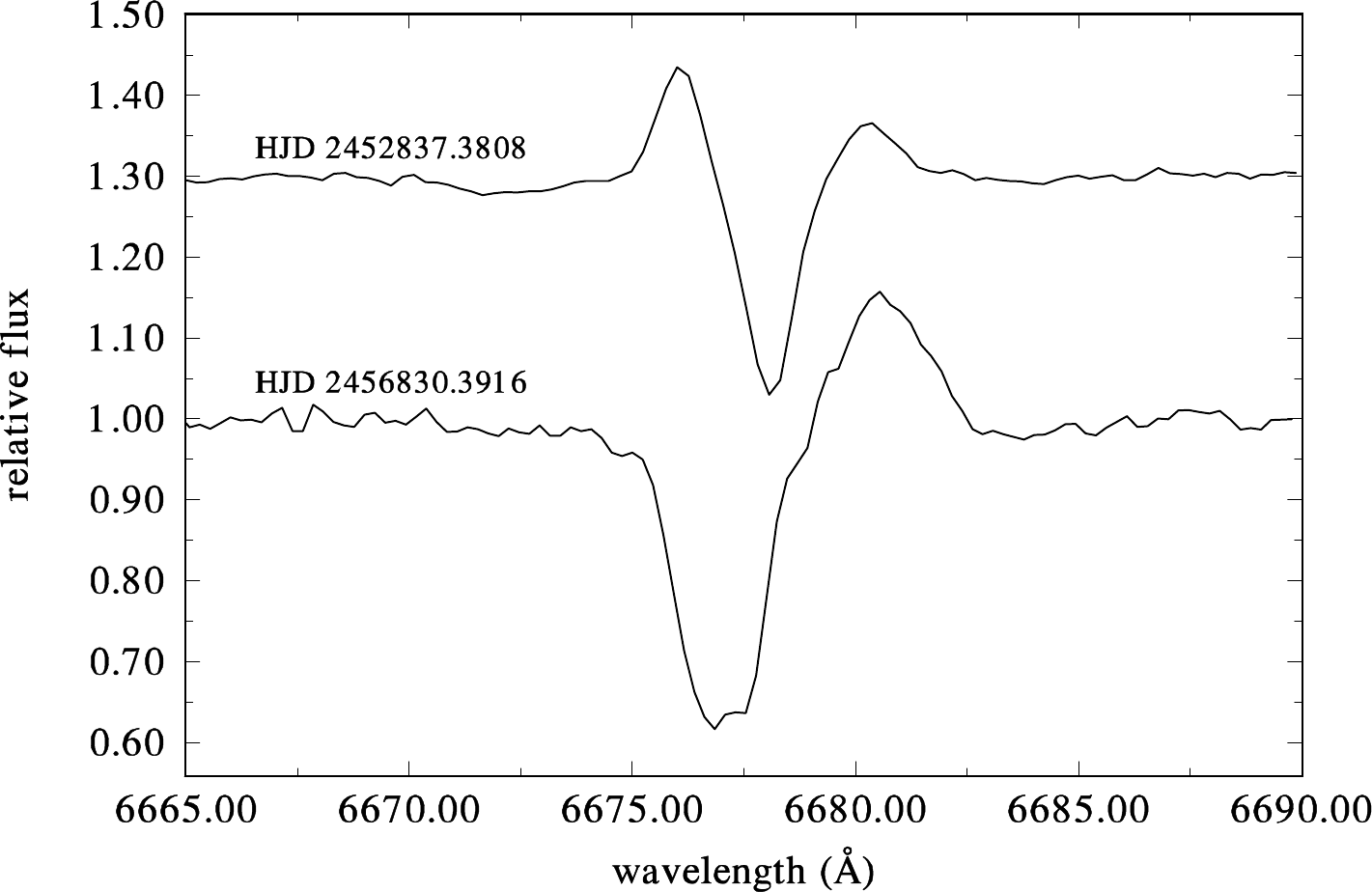}}
\caption{Apparent $V/R$ changes observed for the \ha and \he
emission lines intercompared for two dates. An~apparently anti-phase
behaviour is observed for the later date, when the \he shell RV becomes
quite negative and the He shell line blends with the $V$ peak of the
faint He emission.}
\label{vrdva}
\end{figure}

  To show how confusing the interpretation of $V/R$ changes can be,
we compare the $V/R$ changes of \ha and \ion{He}{i}~6678~\AA\ for
two shorter time intervals in Fig.~\ref{fales}. In the first interval,
the variations for both lines are in phase, while in the second interval,
antiphase variation is observed. We note that this is a consequence
of the fact that the He shell RV becomes very negative (a~temporarily
elongated envelope?) and apparently weakens the $V$ peak of a~relatively
faint He emission. This is best illustrated in Fig.~\ref{vrdva}, where
the \ha and \he profiles for two dates are shown.

This study of \va demonstrates very clearly how hard it is to identify
and understand mutually coexisting and overlapping variability patterns
governing the observed spectral, light, and colour changes. Attempts at modelling them
quantitatively, planned for a~continuation of this study, are expected to shed
more light on the mysterious Be phenomenon. We also suggest that further
monitoring of the object with systematic photometry, high-resolution
spectroscopy, and especially with the optical interferometry could help to
reveal the secrets of this intriguing Be binary, or possibly a multiple system,
as indicated by the analysis of the astrometric data \citep{brandt2021}.

\begin{acknowledgements}
We gratefully acknowledge the use of the latest publicly available version
of the program \fotel written by P.~Hadrava.
We thank A.~Aret, A.~Budovi\v{c}ov\'a, P.~Chadima, M.~Dov\v{c}iak,
J.~Fuchs, P.~Hadrava, J.~Jury\v{s}ek, E.~Kiran, L.~Kotkov\'a,
R.~K\v{r}i\v{c}ek, J.~Libich, J.~Nemravov\'a, P.~Rutsch, S.~Saad, P.~\v{S}koda,
S.~\v{S}tefl, and V.~Votruba, who obtained some of the Ond\v{r}ejov spectra
used in this study. J.R.~Percy kindly put the archive of his systematic
\ubv\ observations of bright Be stars in three observatories at our disposal.
We also acknowledge the constructive suggestions of an anonymous referee
to the first version of this study.
Over the years, this long-term project was supported by the grants 205/06/0304,
205/08/H005, P209/10/0715, and GA15-02112S  of the Czech Science Foundation,
by the grants 678212 and 250015 of the Grant Agency of the Charles University
in Prague, from the research project AV0Z10030501 of the Academy of Sciences
of the Czech Republic, and from the Research Program MSM0021620860
{\sl Physical study of objects and processes in the solar system and
in astrophysics} of the Ministry of Education of the Czech Republic.
The research of PK was supported by the ESA PECS grant 98058.
HB, DR, and DS acknowledge financial support
from the Croatian Science Foundation
under the project 6212 ``Solar and Stellar Variability".
This work has made use of data from
the European Space Agency (ESA) mission Gaia
(\url{https://www.cosmos.esa.int/gaia}), processed by the Gaia
Data Processing and Analysis Consortium (DPAC;
\url{https://www.cosmos.esa.int/web/gaia/dpac/consortium}).
Funding for the DPAC has been provided by national institutions,
in particular the institutions participating in
the Gaia Multilateral Agreement. We also used some spectra
of the BeSS database, operated at LESIA,
Observatoire de Meudon, France: \url{http://basebe.obspm.fr}.
Finally, we acknowledge the use of the electronic database from
the CDS, Strasbourg, and the electronic bibliography maintained by
the NASA/ADS system.
\end{acknowledgements}

\bibliographystyle{aa}
\bibliography{citace}

\begin{appendix}

\section{Details of photometric observations and their reduction and
homogenisation}\label{apb}

\begin{table}
\caption[]{Standard \ubvr\ magnitudes of the stars used
by different observers as comparison stars in their differential
observations of V1294~Aql.}\label{comp1294}
\begin{flushleft}
\begin{tabular}{rcccrcll}
\hline\noalign{\smallskip}
Star&HD&$V$&\bv&\ub&$\vr$\\
    &  &(mag.)&(mag.)&(mag.)&(mag.)\\
\noalign{\smallskip}\hline
\hline\noalign{\smallskip}
V1431 Aql&183324&5.799&0.089&   0.057&0.100\\
  HR 7397&183227&5.844&0.017&$-$0.380&0.087\\
  HR 7438&184663&6.373&0.408&$-$0.033&0.388\\
\noalign{\smallskip}\hline
\end{tabular}
\end{flushleft}
\tablefoot{All tabulated \ubvr\ magnitudes of these stars are based
on numerous all-sky observations secured at Hvar.}
\end{table}

\begin{table*}
\caption[]{\cite{lynds59} versus standard Johnson magnitude
differences for all suitable microvariables observed by Lynds.
The mean \ubv\ values of all considered stars were taken (with one exception)
from the General Catalogue of Photometric Data
\citep[see][]{gcpd}. The yellow magnitude differences `variable minus comparison'
of Lynds are denoted by d$y$.
The \ubv\ magnitudes of the corresponding comparison are always
listed below each program star.}\label{lynds}
\begin{flushleft}
\begin{tabular}{rrcrrrrrrrcll}
\hline\noalign{\smallskip}
Star&HD&$V$   &\bv   &\ub   &d$y$  &d$V$  &\tria(d$V-$d$y$)&\tria(\bv)&Notes\\
    &  &(mag.)&(mag.)&(mag.)&(mag.)&(mag.)&(mag.)   &(mag.)\\
\noalign{\smallskip}\hline
\hline\noalign{\smallskip}
  KP Per  & 21803& 6.405& 0.028&-0.713& 1.100& 1.093& -0.007& -0.373\\
 HR 1182  & 21770& 5.312& 0.401&-0.020&      &      &       &       \\
\noalign{\smallskip}\hline
 $o$ Per  & 23180& 3.833& 0.049&-0.741&-2.827&-2.826& -0.001& -0.027\\
BD+31 649 & 23478& 6.659& 0.076&-0.566&      &      &       &       \\
\noalign{\smallskip}\hline
BD+52 714 & 23675& 6.727& 0.448&-0.542& 0.020&-0.012& -0.032&  0.076&1\\
BD+52 726 & 24431& 6.739& 0.372&-0.610&      &      &       &       \\
\noalign{\smallskip}\hline
BD+52 715 & 23800& 6.922& 0.329&-0.534& 0.212& 0.183& -0.029& -0.043&2\\
BD+52 726 & 24431& 6.739& 0.372&-0.610&      &      &       &       \\
\noalign{\smallskip}\hline
V600 Her  &149881& 7.03 &-0.19 &-0.97 & 0.632& 0.65 &  0.018& -0.103\\
V773 Her  &149822& 6.380&-0.087&-0.188&      &      &       &       \\
\noalign{\smallskip}\hline
V986 Oph  &165174& 6.139&-0.011&-0.928& 1.707& 1.704& -0.003& -0.042\\
NSV10009  &164577& 4.435& 0.031& 0.009&      &      &       &       \\
\noalign{\smallskip}\hline
  ES Vul  &180968& 5.432& 0.038&-0.773&-1.460&-1.475& -0.015& -0.134&2,3\\
BD+21 3719&180889& 6.907& 0.172&0.162\\
\noalign{\smallskip}\hline
V819 Cyg  &188439& 6.293&-0.116&-0.914& 0.058& 0.083&  0.025& -0.474&4\\
 HD 7577  &188074& 6.210& 0.358& 0.032&      &      &       &       \\
\noalign{\smallskip}\hline
V373 Cas  &224151& 6.003& 0.207&-0.713&-1.152&-1.197& -0.045&  0.067&2\\
BD+56 3127&224624& 7.212& 0.14 & 0.08 &      &      &       &       \\
\noalign{\smallskip}\hline
\end{tabular}
\tablefoot{{\sl Notes:} 1... When adopting the mean \ubv\ values for this star,
we omitted the deviating value $V=$6\m88 denoted as uncertain in
the original study by \citet{fernie83}; 2... Be star; 3...The \ubv\ values
for the program and the comparison star were adopted from numerous calibrated
all-sky Hvar \ubv\ observations;
4...In his Fig.~14, \citet{lynds59} erroneously writes
that the magnitude differences shown are in the sense HD~188439 minus
HD~188252. However, HD~188252 was another of his program stars, and
the true comparison for HD~188439 was HD~188074 (cf. his Table~3).
}
\end{flushleft}
\end{table*}

Differential observations from Hvar, Ond\v{r}ejov,
San Pedro Mart\ii r, Tubitak, and \c{C}anakkale were all reduced and carefully
transformed into the standard Johnson \ubv\ system with the help of
the reduction program {\tt HEC22;} see \citet{hhj94} and \citet{hechor98}
for the observational strategy and details of the data reduction.
\footnote{The whole program suite with a detailed manual,
examples of data, auxiliary data files, and results is available at
{\sl http://astro.troja.mff.cuni.cz/ftp/hec/PHOT}\,.}
The latest {\tt HEC22 rel.18} version was used. It allows the use of
non-linear transformation formul\ae, includes the colour extinction coefficient
among the seasonal transformation coefficients, and allows the time
variation of linear extinction coefficients to be modelled in the course of
the observing nights. Our standard strategy is to observe the check star as
frequently as the variable to determine the real data quality. This also permitted replacing the originally selected comparison by the check star when
the former was found to be microvariable after some time. This was indeed
the case of observations of \ve. Our original comparison
35~Aql = HD~183324 = HR~7400
was reported to be a $\lambda$~Boo microvariable with a period of 0\fd0211
and a full amplitude of 0\m02 by \citet{kus94} and received the variable star
name V1431~Aql. We now use our original check star HR~7397 = HD~183227 as
the comparison, but we note that because our observing sequences usually consisted
of three to five cycles of individual observations, they are typically some
0\fd05 -- 0\fd06 long and the variability of 35~Aql is in a sense smeared out.
We therefore retained some series of observations relative to 35~Aql, where
the change to the new comparison HR~7397 would decrease the quality of
differential observations.

Below, we provide some details of the individual data sets defined
in Table~\ref{jouphot} and their reductions.
\begin{itemize}
\item {\sl Station 01 -- Hvar:} \ \
 These differential observations have been secured by a number of observers
over many years, with HR~7397 and HR~7438 as the comparison and check star,
respectively, within the framework of the long-term program of monitoring
the light and colour variations of bright Be stars. The all-sky magnitudes
from Hvar were carefully homogenised and serve as our primary source
of \ubv\ and \ubvr\ magnitudes. The Hvar mean values, which were always added
to magnitude differences from all stations, are listed in
Table~\ref{comp1294}. Since the summer of 2013, the $R$ filter was installed
to the Hvar photometer, and \ubvr\ observations were collected. We note that the transparency curve of the $R$ filter closely corresponds
to that of the standard Cousins $R$ filter. However, because we were
not able to find enough northern bright standard stars with the Cousins \vr\
values, we derived robust mean values of Johnson \vr\ indices from
\citet{john66} and reduced our observations to the Johnson system.
\item {\sl Station 02 -- Brno:} \ \
These \bvr\ observations were secured by P.~Svoboda in his private observatory
with a Sonnar 0.034 m refractor and a CCD camera. They were reduced to the
form of magnitude differences by the author. The Hvar all-sky values of
the comparison star were then added to them.
\item {\sl Station 04 -- Ond\v{r}ejov:} \ \ \ubv\ observations secured with the 0.65~m
Ond\v{r}ejov reflector and a photometer with an EMI tube and transformed into the standard
Johnson system.
\item {\sl Station 12 -- La Silla ESO:} \ \ There are two independent observation sets.
The early observations are all-sky \uvby\ observations published by \citet{kozok85}.
The rest are differential \uvby\ observations obtained with
the Danish 0.50 m reflector (later called Str\"omgren Automatic Telescope (SAT)
\citep[see][for the published data and details of the observations and
reduction]{esovar91,esovar93,esovar94,esovar95} relative to 35~Aql,
HD~7397 being used as the check star. We then used the transformation
from the Str\"omgren into the Johnson system derived by \citet{hecboz01}
to obtain values that can be directly compared to other \ubv\ datasets.
\item {\sl Station 20 -- Toronto} \ \ Differential $BV$ observations secured during
the international campaign on bright Be stars by John Percy, who kindly placed them at
our disposal. They were transformed into the standard Johnson system via linear
transformations by the author, and the Hvar all-sky values of the comparison star
were added to the magnitude differences.
\item {\sl Station 26 -- Haute Provence} \ \ Two \ubv observations were secured by
\citet{haupt74}. Individual observations were kindly provided by H. Haupt upon our request
and corrected to our adopted values for the comparison star.
\item {\sl Station 30: SPM 1.50~m and 0.84~m reflectors} \ \  The early all-sky observations
(before JD~2446000) were secured at the 1.50~m reflector in the 13C system and reduced by
their authors \citep{alvschu82,schugui84}. They were then transformed into the standard
Johnson \ubv\ system with the help of transformation formul\ae\  derived by \citet{hecboz01}.
More recent differential \ubv\ observations were secured by MW with the 0.84~m reflector
and Cuenta-pulsos photon-counting photometer. Variable extinction during the nights
was monitored, and the data were reduced to the standard \ubv\ system.
\item {\sl Station 37 -- Jungfraujoch} \ \ These all-sky seven-colour (7-C)
observations were secured in the Geneva photometric system at the Jungfraujoch Sphinx
mountain station 0.40 m reflector and kindly placed at our disposal by G.~Burki. They were transformed
into the standard Johnson \ubv\ system with the help of transformations derived
by \citet{hecboz01}.
\item {\sl Station 44: Mt Palomar 0.51 m reflector} \ \
\citet{lynds59} carried out numerous differential observations of
early-type stars classified as giants, including \ve. He observed
in yellow light, with a Corning~3384 filter and an EMI~6094 tube.
Most of his observations are only published in the form of time plots
of the data, variable minus comparison. In Table~\ref{lynds} we
compare the mean magnitude differences between the respective stars
for all his microvariables with the mean differences in the standard
Johnson $V$ filter. It is obvious that for all practical purposes,
the yellow filter used by Lynds measures magnitudes that are very close
to the Johnson $V$ filter. The largest \tria(d$V-$d$y$)
difference was found for V373~Cas, which is now known to be a Be star
and a spectroscopic binary \citep{373cas}. The Hipparcos \hp\ magnitude
has a 0\m1 range of variations \citep{esa97}.
We therefore digitised the yellow observations of \va
from the enlarged Fig.~13 of \citet{lynds59} and derived HJDs
and magnitude differences that were then added to the $V$ magnitude
of the comparison star HR~7438 from Table~\ref{comp1294}. We estimate
that the HJDs are accurate to $\pm$0\D0003, which is very satisfactory.
\item {\sl Station 61 -- Hipparcos:} \ \ These all-sky observations were
reduced to the standard $V$ magnitude via the transformation formul\ae\
derived by \citet{hpvb}. To derive this transformation, the correct values of the Johnson \bv\ and \ub\ indices are required.
We checked the overlapping Hvar \ubv\ photometry to find that the \bv\
index remained basically constant over the time interval covered by
the \hp\ photometry at a value of 0\m10. The \ub\ index varied from
$-$0\m77 to $-$0\m64, but we verified that these extremes cause an error of only
about 0\m01 in the transformation into Johnson $V$. We therefore used
\bv$=-$0\m10 and \ub$=-$0\m70 to transform \hp\ into $V$.
\item {\sl Station 66 -- Tubitak:} \ \ These differential \ubv\ observations
were secured with the 0.40~m reflector of the Turkish national observatory
and an~SSP-5A solid-state photometer and were transformed into the standard system.
\item {\sl Station 89 -- \c{C}anakkale:} \ \ These differential \ubv\ observations
were secured at \c{C}anakkale mountain station with a 0.40~m reflector and an SSP-5
solid-state photometer and were transformed into the standard system.
\item {\sl Station 93 -- ASAS3 V photometry:} \ \ We extracted these
all-sky observations from the ASAS3 public archive \citep{pojm2002},
using the data for diaphragm~1, which has the lowest rms errors on average.
We omitted all observations of grade D and observations with rms errors
larger than 0\m04. We also omitted a strongly deviating observation at
HJD~2452662.6863.
\end{itemize}
\end{appendix}
\end{document}